\documentclass[10pt,journal,final]{IEEEtran}
\IEEEoverridecommandlockouts

\usepackage{amsfonts,amssymb}
\usepackage{ifpdf}
\usepackage{cite}
\usepackage{multirow}
\usepackage[hyphens]{url}
\usepackage{stfloats}
\usepackage{bm}
\usepackage{color,xcolor}
\usepackage{subfigure}
\usepackage{caption}

\ifCLASSINFOpdf
   \usepackage[pdftex]{graphicx}
   \graphicspath{{../pdf/}{../jpeg/}}
   \DeclareGraphicsExtensions{.pdf,.jpeg,.png}
\else
   \usepackage[dvips]{graphicx}

   \graphicspath{{../eps/}}

   \DeclareGraphicsExtensions{.eps}
\fi

\usepackage[cmex10]{amsmath}

\usepackage{algorithm}
\usepackage{algorithmic}
\ifCLASSOPTIONcompsoc
\else
  \usepackage[caption=false,font=footnotesize]{subfig}
\fi
\usepackage{makecell}
\begin{document}


\title{Implementing Neural Networks Over-the-Air  via Reconfigurable Intelligent Surfaces}
\author{Meng Hua,~\IEEEmembership{Member,~IEEE,}
Chenghong~Bian,~\IEEEmembership{Student Member,~IEEE},
Haotian~Wu,~\IEEEmembership{Student Member,~IEEE}, and 
Deniz~G\"und\"uz,~\IEEEmembership{Fellow,~IEEE}
\thanks{This work was  supported by the SNS JU Project 6G-GOALS under the	EU’s Horizon Program with Grant 101139232.
}
\thanks{The authors are with the Department of Electrical and Electronic Engineering, Imperial College London, London SW7 2AZ, U.K. (e-mail: \{m.hua,c.bian22,haotian.wu17,d.gunduz\}@imperial.ac.uk).
}

}
\vspace{-1.5cm}
\maketitle
\begin{abstract}
By leveraging the  superposition property, over-the-air computation (OAC) of waveforms enables  computations to be performed in an analog fashion in wireless environments, leading to faster computation, lower latency, and reduced energy consumption. In this paper, we  investigate reconfigurable intelligent surface (RIS)-aided multiple-input-multiple-output (MIMO) OAC systems designed to emulate the  fully-connected (FC) layer of a neural network (NN) via analog OAC, where the RIS and the transceivers are jointly adjusted to engineer the ambient wireless propagation environment to 
emulate the weights of the target FC layer. We refer to this novel computational paradigm as \textit{AirFC}.
We first study the case in which  the precoder, combiner, and  RIS phase shift matrices are jointly optimized to minimize the mismatch between the OAC system and the target FC layer. To solve this non-convex optimization problem, we propose a low-complexity alternating optimization algorithm, where semi-closed-form/closed-form solutions for all optimization variables are derived. 
Next, we consider training of the system parameters using two distinct learning strategies,  namely  \textit{ centralized training} and  \textit{distributed training}. 
In the centralized training approach, training is performed at either the transmitter or the receiver, whichever possesses the channel state information (CSI), and the trained parameters are provided to  the other terminal.
In the distributed training approach,  the transmitter and receiver  iteratively update their parameters through back and forth  transmissions by leveraging  channel reciprocity, thereby avoiding CSI acquisition and significantly reducing   computational complexity.
Subsequently, we extend our analysis to a multi-RIS scenario by exploiting its spatial diversity gain to enhance the system performance, i.e., classification accuracy. Simulation results show that the AirFC system realized by the RIS-aided MIMO configuration achieves satisfactory classification accuracy. Furthermore, it is shown that the multi-RIS system brings significant improvement  in terms of the classification accuracy, especially 
in line-of-sight (LoS)-dominated wireless environments.
\end{abstract}
\begin{IEEEkeywords}
Over-the-air computation,  analog computation, reconfigurable intelligent surface, neural networks,  programmable wireless environment
\end{IEEEkeywords}

\section{Introduction}
We are witnessing a rapid growth of data traffic due to the increasing popularity of edge devices. Integrated  with artificial intelligence, these devices enable ubiquitous computing and ambient intelligence  \cite{Letaief20196roadmap}. 
However, processing  massive amounts of data at the edge devices presents significant challenges due to their limited computational resources and battery-powered energy constraints. To address these challenges, offloading computation or inference tasks from edge devices to resource-rich edge servers via wireless channels is gaining increasing attention \cite{Gunduz2020communicate,mao2017survey,cao2020overview,chen2019deep}. 
In the conventional \textit{transmit-then-compute} approach,  the edge device transmits signals to the edge server via the wireless channel, where the server decodes and processes the signals. However, this method often suffers from excessive latency and poor spectrum utilization. 
A promising alternative,  termed  \textit{over-the-air computation (OAC)}, exploits the superposition property of the wireless multiple access channel to perform simple algebraic operations, such as weighted sums, arithmetic means, Euclidean norms, etc., inherently combining computation with communication \cite{4Goldenbaum2009function,sahin2023survey,zhao2023Intelligent,liu2021over,zhu20196mimo}.  Thanks to its spectral and energy efficiency, OAC has been employed in distributed training \cite{Amiri2020federated,mohammadi2020machine} as well as inference tasks \cite{yilmaz2025private,liu2024over}.


Typically,  machine learning (ML) models are trained in the digital domain using graphics-processing units, e.g., \cite{russakovsky2015imagenet,wu2023deep,peng2019modulation}, which are time-consuming and energy-intensive due to the separation of memory and processing inherent in von Neumann architectures. To address this limitation, one promising solution is to train  neural networks (NNs) using physical hardware based on optics, mechanics, and electronics, known as physical neural networks (PNNs), operating in the analog domain  \cite{wright2022deep,van2023retrainable,gao2025disaggregated}. This approach builds deep PNNs composed of layers of controllable physical systems, where the computation traditionally performed in the digital domain is approximately replaced by the transmission of waves or optics through physical media. Since the computation naturally occurs over the air, with signal velocities approaching the speed of light, training PNNs can significantly reduce both latency and power consumption. Inspired by this idea,  MIMO OAC  has been explored for realizing fully-connected (FC) layers with fast and energy-efficient computations due to their structural similarities. 
On one hand, both the MIMO OAC paradigm and  FC layers in digital NNs resemble point-to-point MIMO systems.  On the other hand,  both primarily involve simple algebraic operations such as multiplication and addition. 
To date, there has been only limited work investigating the use of the OAC paradigm to implement NN functions \cite{hughes2019wave, kendall2020training,reus2023airfc,yang2023over}.   For instance, \cite{hughes2019wave}  identified a mapping between wave physics dynamics and a prototype recurrent NN, demonstrating that acoustic signals, by leveraging wave physics properties, can perform vowel classification on raw audio signals. Similarly, \cite{kendall2020training}  applied the stochastic gradient descent method to train an analog NN by adjusting the conductance of programmable resistive diodes.  In \cite{reus2023airfc}, a linear function in the FC layer was realized via the superposition of orthogonal frequency-division multiplexing (OFDM) signals emitted by multiple transmitters and received at a single receiver, i.e., a multiple-input single-output (MISO) system. This was subsequently extended to a MIMO system in \cite{yang2023over}, where the precoder and combiner were jointly trained over the air, achieving satisfactory classification accuracy. However, in the aforementioned works, the waveform design must adapt to the channel, indicating that system performance is heavily influenced by channel conditions, such as fading and channel rank.

The emerging technology of reconfigurable intelligent surface (RISs) offers a new paradigm for wireless communications \cite{pan2021Reconfigurable,mei2022Intelligent,chen2022active,meng2022Intelligent,chen2025intelligent,wu2019intelligentxx}, attracting considerable attention and being considered a candidate technology for sixth-generation (6G) networks. An RIS is a planar metasurface composed of massive low-cost passive reflecting elements, such as PIN diodes and varactor diodes, each of which can be digitally controlled to independently adjust the amplitude and/or phase shift of incident signals, thereby reshaping wireless signal propagation. 
Due to this appealing feature, RIS has been extensively studied and incorporated into various wireless systems, such as mobile edge computing \cite{chen2023RISMEC}, sensing \cite{Hua2024Intelligent}, and integrated sensing and communication  \cite{Meng2024sensing}. Naturally, RIS is also appealing for analog computation \cite{liu2022programmable,chen2024RIS,Garcia2023irNN,zhang2024radio,liu2025over}.  For instance, \cite{liu2022programmable}   presented a multi-layer RIS experimental prototype that accomplished a range of deep learning tasks, including image classification, mobile communication coding-decoding, and real-time multi-beam focusing.  Subsequently, RIS-based over-the-air semantic communication was studied in \cite{chen2024RIS}. 
In \cite{Garcia2023irNN}, the authors proposed utilizing RISs to realize 1D convolution operations by exploiting multipath delays, where each channel impulse response acts as an individual finite impulse response filter, convolving with the transmitted signal to emulate a digital convolutional NN.  Building upon this idea, the work  in \cite{zhang2024radio} further extended the concept to 2D convolution, enabling RISs to emulate more complex convolutional neural network structures for spatial signal processing.   The authors of \cite{liu2025over} designed a complex-valued neural network inspired by ordinary differential equations (ODEs), termed the Air-ODE network, which uses the physics of RIS-based signal reflections to execute computations during transmission.
However, these prior works focus either  on single-input single-output (SISO) or multiple-input and single-output (MISO)  systems. The extension to  MIMO systems, particularly the principle of combining MIMO architectures with RISs to emulate digital layers, remains unexplored.

In this paper, we study RIS-aided MIMO  OAC  to replicate the functionality of digital FC layers, in which the RIS and transceivers are jointly adjusted to engineer the ambient wireless propagation environment to emulate the weight matrix of a digital FC layer. We term this novel computational paradigm \textit{AirFC}.
In contrast to   \cite{liu2022programmable,chen2024RIS,Garcia2023irNN,zhang2024radio,liu2025over}, our work extends the concept from SISO and MISO systems to  MIMO systems and investigates the joint design of the precoder, the combiner, and the RIS phase shift to emulate digital FC layers. Note that although work \cite{yangyuzhi2024realizing} studied an OAC NN with RIS-aided MIMO systems, it primarily  focused on constructing NNs through iterative RIS-aided transmissions and approximate over-the-air gradient backpropagation, rather than accurately emulating digital fully connected layers.  Another closely related work is \cite{stylianopoulos2025over}, which explores the integration of RIS-assisted MIMO systems into edge inference by treating the wireless propagation environment as a hidden layer of a NN. In contrast, our proposed AirFC framework aims to precisely emulate digital FC layers through analog OAC, leveraging  RIS-assisted MIMO systems where the RIS phase shifts, precoder, and combiner matrices are jointly optimized or  trained  to emulate digital FC layers.
Particularly,  we analyze the impact of multi-RIS configurations on channel conditions and evaluate the corresponding ML system performance.   
The main contributions of this paper are summarized as follows.

\begin{itemize}
    \item First, we consider the case in which  the precoder, the combiner, and the RIS phase shift matrices, are optimized at each channel state to emulate a FC layer of a target NN over-the-air. 
    To solve this non-convex optimization problem, we propose an efficient alternating optimization algorithm by partitioning the optimization variables into three blocks and optimizing each block alternately until convergence. In particular, a closed-form solution for the precoder is derived based on the Lagrange duality method, and semi-closed-form solutions for the combiner and RIS phase shifts are provided, resulting in an extremely low-complexity algorithm.
    \item Second, instead of emulating a target NN, the system parameters are trained using a customized loss function that accounts for practical transmit power constraints. Two learning strategies are proposed: \textit{centralized training} and \textit{distributed training}.
    In the centralized training approach, training is performed at either the transmitter or the receiver, whichever possesses the channel state information (CSI), and the corresponding trained parameters are provided to  the other terminal.
    We also consider an alternative over-the-air training method, namely distributed training, which does not require CSI. In this approach, the transmitter and receiver exchange information by leveraging channel reciprocity, and iteratively update parameters through back and forth  transmissions. By leveraging gradient-based updates for the precoder and combiner, the proposed method achieves significantly lower computational complexity than its centralized counterpart.
    
    \item Third, we extend our model to the multi-RIS scenario. We theoretically analyze and demonstrate the performance improvement achieved by deploying multiple RISs compared to a single RIS.
    \item Finally, simulation results validate that the RIS-aided MIMO OAC system can effectively emulate the  FC NN layers. The proposed over-the-air ML training strategies also achieve satisfactory classification accuracy. Moreover, the multi-RIS system significantly outperforms the single-RIS setup in terms of classification accuracy, especially 
    in line-of-sight (LoS)-dominated wireless environments.	

\end{itemize}
The rest of this paper is organized as follows. Section II introduces two paradigms: RIS-aided OAC for a given FC layer and a novel trainable over-the-air transmission architecture. Section III proposes an algorithm to realize AirFC emulation. Section IV studies the RIS-aided over-the-air ML implementation. Section V extends the single-RIS case to the multi-RIS scenario. Section VI provides numerical results. Finally, the paper is concluded in Section VII.

\emph{Notations}: Boldface lower-case and upper-case letters denote vectors and matrices, respectively.  ${\mathbb C}^ {d_1\times d_2}$  represents the set of  complex-valued $d_1\times d_2$  matrices. For a complex-valued vector $\bf x$, ${\left\| {\bf x} \right\|}$ represents the  Euclidean norm of $\bf x$, ${\rm arg}({\bf x})$ denotes the phase of   $\bf x$, ${{\bf{x}}^*}$ denotes its conjugate, and ${\rm diag}(\bf x) $ denotes a diagonal matrix whose main diagonal elements are extracted from vector $\bf x$.
For a square matrix $\bf X$, ${\bf X}^*$, ${\bf X}^H$, ${\rm{Tr}}\left( {\bf{X}} \right)$,  ${{\bf{X}}^{ - 1}}$,   ${\rm{rank}}\left( {\bf{X}} \right)$, and ${\left\| {\bf{X}} \right\|_F}$ represent its conjugate, conjugate transpose, trace,  inverse, rank, and Frobenius norm, respectively.   ${ {\bf{X}} _{i,i}}$ represents the $i$th diagonal element of   matrix $\bf X$. A circularly symmetric complex Gaussian (CSCG) random variable $x$ with mean $ \mu$ and variance  $ \sigma^2$ is denoted as ${x} \sim {\cal CN}\left( {{{\mu }},{{\sigma^2 }}} \right)$. We denote the statistical expectation operation by ${\mathbb E}\left\{  \cdot  \right\}$,       Hadamard product notation by $ \odot $, and use ${\cal O}\left(  \cdot  \right)$ for the big-O computational complexity notation.

\section{System Model}

\begin{figure}[!t]
	\centerline{\includegraphics[width=3in]{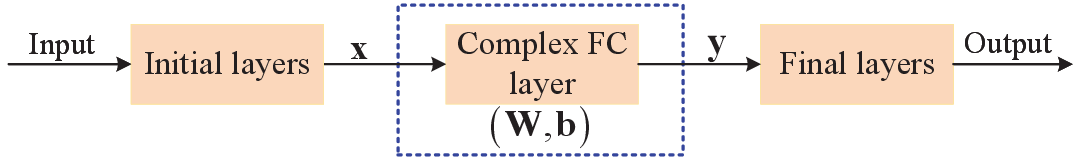}}
	\caption{Target  NN architecture.} \label{model_fig1}
\end{figure}

\begin{figure}[!t]
	\centerline{\includegraphics[width=3.5in]{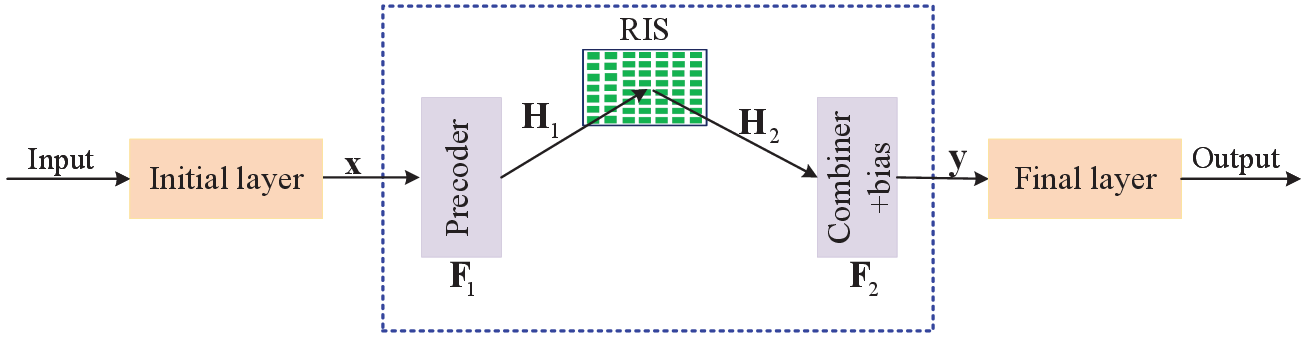}}
	\caption{Analog transmission layer implementation of the target NN.} \label{model_fig1_1}
\end{figure}
We consider a target  NN architecture as illustrated in Fig.~\ref{model_fig1}, divided into three components, namely initial layers, a middle layer, and final layers. The middle layer is a  complex FC layer, while the initial and final layers may contain large numbers of functional layers, such as residual neural network,  transformer layer, multi-head attention layer,  depending on specific applications. Let ${\bf{x}} \in {{\mathbb C}^{N \times 1}}$  and ${\bf{y}} \in {{\mathbb C}^{N \times 1}}$ represent the input and output of the complex FC layer, respectively, where $N$ denotes the input/output dimension of the FC layer. The relationship between the  input and output of the complex middle layer is given by 
\begin{align}
	{\bf{y}} = {\bf{Wx}} + {\bf{b}}, \label{linearfunction}
\end{align}
where  ${\bf{W}} \in {{\mathbb C}^{N \times N}}$ and ${\bf{b}} \in {{\mathbb C}^{N \times 1}}$ denote the weight matrix and bias vector, respectively. 

Our goal is to implement the middle layer of the target NN over-the-air in an analog fashion as shown in Fig.~\ref{model_fig1_1}. In the analog implementation, the middle layer is replaced by the so-called ``transmission layer" consisting of  three components, namely the precoder, the RIS, and the combiner.  It is assumed that the direct link between the transmitter and receiver is blocked for clarity, although the existence of a direct link can be easily incorporated into our model. Throughout the paper, we assume perfect CSI is available whenever CSI is required for training or optimization purposes.

In this paper, we will investigate two cases. In the first case,  parameters of the  transmission layer are non-trainable and are designed to approximate a  given target FC layer. In the second case,  over-the-air parameters are trainable and are updated according to a loss function, similarly to the weight update mechanism in NN training.
\subsection{RIS-aided OAC to Emulate a Target   FC Layer} \label{subsection:OAC}
In this subsection, we study the case with non-trainable over-the-air parameters. We assume that the number of transmit and receive antennas is  $N$, matching the input/output dimension of the complex FC layer depicted in Fig.~\ref{model_fig1}.  Let $M$ denote the number of RIS reflecting elements.   Denote by ${\bf H}_1\in {{\mathbb C}^{M \times N}}$ and ${\bf H}_2\in {{\mathbb C}^{N \times M}}$ the complex equivalent baseband channels between the transmitter and the RIS, and between the RIS and the receiver, respectively. As shown in Fig.~\ref{model_fig1_1}, the received signal at the receiver, combined with the bias, is given by
\begin{align}
{\bf{y}} = {{\bf{F}}_2}\left( {{{\bf{H}}_2}{\bf{\Theta }}{{\bf{H}}_1}{{\bf{F}}_1}{\bf{x}} + {\bf{n}}} \right) + {\bf{b}}, \label{channel:singleRIS}
\end{align} 
where ${{{\bf{F}}_1}}\in {{\mathbb C}^{N \times N}}$ and ${{{\bf{F}}_2}}\in {{\mathbb C}^{N \times N}}$ denote the transmit precoder and the receive combiner, respectively, ${\bf{\Theta }} = {\rm{diag}}\left( {{e^{j{\theta _1}}},{e^{j{\theta _2}}}, \ldots ,{e^{j{\theta _M}}}} \right)$ represents the diagonal reflection coefficient matrix of the RIS, with $\theta_m, m \in \left\{ {1,2 \ldots ,M} \right\}$, denoting the $m$th  phase shift.  In addition, ${\bf{n}} \sim {\cal CN}\left( {0,{\sigma ^2{\bf I}_N}} \right)$ denotes the additive white Gaussian noise,  and ${\bf{b}}$ is the bias as defined in \eqref{linearfunction}.

To apply over-the-air computation to emulate the linear complex FC layer,  we formulate an emulation error minimization problem by comparing \eqref{linearfunction} and \eqref{channel:singleRIS}, as follows:
\begin{subequations}  \label{P1}
\begin{align}
    &\mathop {\min }\limits_{{{\bf{F}}_1},{{\bf{F}}_2},{\bf{\Theta }}} \left\| {{{\bf{F}}_2}{{\bf{H}}_2}{\bf{\Theta }}{{\bf{H}}_1}{{\bf{F}}_1} - {\bf{W}}} \right\|_F^2 + {{\mathbb E}_{\bf{n}}}\left\{ {{{\left\| {{{\bf{F}}_2}{\bf{n}}} \right\|}^2}} \right\} \label{objectivefunction}\\
   & {\rm{s}}{\rm{.t}}{\rm{. }}~\left\| {{{\bf{F}}_1}} \right\|_F^2 \le {P_{{\rm{max}}}},\label{transmitpower}\\
   &\qquad \left| {{{\bf{\Theta }}_{i,i}}} \right| = 1,{\kern 1pt} {\kern 1pt} {\kern 1pt} {\kern 1pt} {\kern 1pt} {\kern 1pt} {\kern 1pt} i = 1, \ldots ,M.\label{RIS_phaseshift}
\end{align}
\end{subequations}
The first term in \eqref{objectivefunction} aims to minimize the emulation error in the weights, while the second term in \eqref{objectivefunction} aims to minimize the bias noise. Constraint \eqref{transmitpower} imposes a power budget lemulation on the transmitter, and \eqref{RIS_phaseshift}  enforces the unit-modulus constraint on each RIS phase shift. 

Problem \eqref{P1} is inherently non-convex since the optimization variables
are highly coupled in the objective function \eqref{objectivefunction}, and a unit-modulus constraint is imposed on  RIS phase shifts. There are no standard methods for solving this type of problem optimally. A low-complexity alternating optimization algorithm is proposed in Section \ref{section:non-trainable}.
\subsection{A Novel Trainable Over-the-Air Transmission Architecture}
In this subsection, we introduce the case with trainable over-the-air parameters. 
 Unlike in Section \ref{subsection:OAC}, where the over-the-air parameters are optimized based on a target FC layer, here we  treat the over-the-air parameters as trainable weights. 
We consider a modified loss function consisting of a cross-entropy loss term,  assuming that the underlying task is a classification task,
and a penalty term associated with the transmit power budget constraint. The proposed loss function for training the NN is given by
\begin{align}
{{\cal L}_{{\rm{loss}}}} =  - \sum\limits_{i = 1}^C {{p_i}\log \left( {{{\hat p}_i}} \right)}  - {\lambda _{\rm{p}}}\min \left\{ {0,{P_{\max }} - {P_{{\rm{Tx}}}}} \right\}, \label{lossfunction}
\end{align}
where $C$ denotes the number of classes, ${\lambda _{\rm{p}}}\ge 0$ is the penalty factor,  ${{p_i}}$ and ${{{\hat p}_i}}$ denote the true label and predicted softmax probability of the $i$th class, respectively, and ${{P_{{\rm{Tx}}}}}$ denotes the transmit power.  Two  distinct training strategies will be proposed, and the detailed implementations will be presented in Section \ref{section:trainable}.

\section{Non-trainable RIS-aided FC Layer Implementation}\label{section:non-trainable}
In this section, we propose an efficient algorithm to solve problem \eqref{P1} by applying the alternating optimization method, where we divide
all the optimization variables into three blocks, namely ${{{\bf{F}}_1}}$, ${{{\bf{F}}_2}}$, and ${\bf{\Theta }}$, and  alternately optimize each block until convergence is achieved.
\subsubsection{For any given RIS phase shift ${\bf{\Theta }}$ and receive combiner ${{{\bf{F}}_2}}$, the subproblem corresponding to the transmit precoder optimization is given by (ignoring  irrelevant constants)}
\begin{subequations}  \label{P1_transmitbeamformer}
	\begin{align}
	&\mathop {\min }\limits_{{{\bf{F}}_1}} \left\| {{{\bf{F}}_2}{{\bf{H}}_2}{\bf{\Theta }}{{\bf{H}}_1}{{\bf{F}}_1} - {\bf{W}}} \right\|_F^2 \label{transmitbeamformer_objectivefunction}\\
	& {\rm{s}}{\rm{.t}}{\rm{. }}~\eqref{transmitpower}.
	\end{align}
\end{subequations}
This is a quadratically
constrained quadratic program (QCQP). In the following, 
we derive a semi-closed-form yet optimal solution for problem \eqref{P1_transmitbeamformer}  by using the Lagrange duality method \cite{boyd2004convex}. To  be specific, by introducing dual variable $\lambda \ge 0 $ with constraint \eqref{transmitpower},  the Lagrangian function of
problem  \eqref{P1_transmitbeamformer} is given by 
\begin{align}
{\cal L}\left( {{{\bf{F}}_1},\lambda } \right) = \left\| {{{\bf{F}}_2}{{\bf{H}}_2}{\bf{\Theta }}{{\bf{H}}_1}{{\bf{F}}_1} - {\bf{W}}} \right\|_F^2 + \lambda \left( {\left\| {{{\bf{F}}_1}} \right\|_F^2 - {P_{{\rm{max}}}}} \right).
\end{align}
By taking the  first-order derivative of ${\cal L}\left( {{{\bf{F}}_1},\lambda } \right)$ with ${\bf{F}}_1^*$ and setting it to zero, we obtain the optimal solution of ${{{\bf{F}}_1}}$ as 
 \begin{align}
 {\bf{F}}_1^{{\rm{opt}}} = {\left( {{{\bm \Upsilon} ^H}{\bm \Upsilon}   + \lambda {{\bf{I}}_N}} \right)^{ - 1}}{{\bm \Upsilon}  ^H}{\bf{W}}, \label{beamformer}
 \end{align}
where ${\bm \Upsilon}   = {{\bf{F}}_2}{{\bf{H}}_2}{\bf{\Theta }}{{\bf{H}}_1}$. According to the complementary slackness condition \cite{boyd2004convex}, the optimal solution of $ {\bf{F}}_1^{{\rm{opt}}}$ and ${\lambda ^{{\rm{opt}}}}$ should satisfy the following equation:
 \begin{align}
 {\lambda ^{{\rm{opt}}}}\left( {\left\| {{\bf{F}}_1^{{\rm{opt}}}} \right\|_F^2 - {P_{{\rm{max}}}}} \right) = 0.
 \end{align}
There are two possible solutions: 1) ${\lambda ^{{\rm{opt}}}}=0$ and ${\left\| {{\bf{F}}_1^{{\rm{opt}}}} \right\|_F^2 \le {P_{{\rm{max}}}}}$; 2) ${\lambda ^{{\rm{opt}}}}>0$ and ${\left\| {{\bf{F}}_1^{{\rm{opt}}}} \right\|_F^2 ={P_{{\rm{max}}}}}$. To this end, we first check whether  ${\lambda ^{{\rm{opt}}}}=0$ is the optimal solution or not. If 
\begin{align}
\!\left\| {{\bf{F}}_1} \right\|_F^2 - {P_{{\rm{max}}}} = \left\| {{{\left( {{{\bm \Upsilon}  ^H}{\bm \Upsilon} } \right)}^{ - 1}}{{\bm \Upsilon} ^H}{\bf{W}}} \right\|_F^2 - {P_{{\rm{max}}}} < 0,
\end{align}
which indicates that  ${\bf{F}}_1^{{\rm{opt}}} = {\left( {{{\bm \Upsilon}^H}{\bm \Upsilon} } \right)^{ - 1}}{{\bm \Upsilon} ^H}{\bf{W}}$; otherwise, we should calculate $\lambda $ that renders $\left\| {{{\left( {{{\bm \Upsilon} ^H}{\bm \Upsilon}  + \lambda {{\bf{I}}_N}} \right)}^{ - 1}}{{\bm \Upsilon} ^H}{\bf{W}}} \right\|_F^2 = {P_{{\rm{max}}}}$. It can be readily checked that ${{\bm \Upsilon} ^H}{\bm \Upsilon}$ is  a positive semi-definite matrix, we can perform eigendecomposition as 
\begin{align}
{{\bm \Upsilon} ^H}{\bm \Upsilon}  = {\bf{U\Sigma }}{{\bf{U}}^H}, \label{eigendecomposition}
\end{align}
where ${\bf{U}}{{\bf{U}}^H} = {{\bf{U}}^H}{\bf{U}} = {{\bf{I}}_N}$, and ${\bf{\Sigma }}$ is a diagonal matrix consisting of eigenvalues.
 Substituting \eqref{eigendecomposition} into $\left\| {{\bf{F}}_1^{{\rm{opt}}}} \right\|_F^2$ in \eqref{beamformer},  we have 
 \begin{align}
 \left\| {{\bf{F}}_1^{{\rm{opt}}}} \right\|_F^2 &= {\rm{tr}}\left( {{{\left( {{\bf{\Sigma }} + \lambda {{\bf{I}}_N}} \right)}^{ - 2}}{{\bf{U}}^H}{{\bm \Upsilon} ^H}{\bf{W}}{{\bf{W}}^H}{\bm \Upsilon} {\bf{U}}} \right)\notag\\
& = \sum\limits_{i = 1}^N {\frac{{{{\left( {{{\bf{U}}^H}{{\bm \Upsilon}  ^H}{\bf{W}}{{\bf{W}}^H}{\bm \Upsilon}  {\bf{U}}} \right)}_{i,i}}}}{{{{\left( {{{\bf{\Sigma }}_{i,i}} + \lambda } \right)}^2}}}}.
 \end{align}
 It is not difficult to verify that $\left\| {{\bf{F}}_1^{{\rm{opt}}}} \right\|_F^2$ is monotonically decreasing with $\lambda $, and a simple bisection method can be applied for solving it. To confine the search range to $\left[ {0,{\lambda ^{{\rm{up}}}}} \right]$, an upper bound  can be set as: 
 \begin{align}
 {\lambda ^{{\rm{up}}}} = \sqrt {\frac{{\sum\limits_{i = 1}^N {{{\left( {{{\bf{U}}^H}{{\bm \Upsilon}  ^H}{\bf{W}}{{\bf{W}}^H}{\bm \Upsilon}  {\bf{U}}} \right)}_{i,i}}} }}{{{P_{{\rm{max}}}}}}}. 
 \end{align}

\subsubsection{For any given RIS phase shift ${\bf{\Theta }}$ and transmit precoder  ${{{\bf{F}}_1}}$, the subproblem corresponding to the receive combiner optimization is given by (ignoring  irrelevant constants)}

\begin{subequations}  \label{P1_receivecombiner}
	\begin{align}
	&\mathop {\min }\limits_{{{\bf{F}}_2}} \left\| {{{\bf{F}}_2}{{\bf{H}}_2}{\bf{\Theta }}{{\bf{H}}_1}{{\bf{F}}_1} - {\bf{W}}} \right\|_F^2 + {{\mathbb E}_{\bf{n}}}\left\{ {{{\left\| {{{\bf{F}}_2}{\bf{n}}} \right\|}^2}} \right\} \label{receivecombiner_objectivefunction}
	\end{align}
\end{subequations}
Noting that ${{\mathbb E}_{\bf{n}}}\left\{ {{{\left\| {{{\bf{F}}_2}{\bf{n}}} \right\|}^2}} \right\} = {\sigma ^2}{\rm{tr}}\left( {{{\bf{F}}_2}{\bf{F}}_2^H} \right)$, objective function  \eqref{receivecombiner_objectivefunction} is  quadratic and its global solution can be attained. Specifically,  taking the  first-order derivative of $\left\| {{{\bf{F}}_2}{{\bf{H}}_2}{\bf{\Theta }}{{\bf{H}}_1}{{\bf{F}}_1} - {\bf{W}}} \right\|_F^2 + {\sigma ^2}{\rm{tr}}\left( {{{\bf{F}}_2}{\bf{F}}_2^H} \right)$ with ${\bf{F}}_2^*$ and setting it to zero, its optimal solution of ${{{\bf{F}}_2}}$ is given by 
\begin{align}
{{\bf{F}}_2^{\rm opt}} = {\left( {\bar {\bm \Upsilon} {{\bar {\bm \Upsilon} }^H} + {\sigma ^2}{{\bf{I}}_N}} \right)^{ - 1}}{\bf{W}}{{\bar {\bm \Upsilon} }^H}, \label{combiner:optimal}
\end{align}
where $\bar {\bm \Upsilon} = {{\bf{H}}_2}{\bf{\Theta }}{{\bf{H}}_1}{\bf{F}}_1$.

\subsubsection{For any given transmit precoder  ${{{\bf{F}}_1}}$ and receive combiner ${{{\bf{F}}_2}}$, the subproblem corresponding to the RIS phase shift ${\bf{\Theta }}$  optimization is given by (ignoring  irrelevant constants)}
\begin{subequations}  \label{P1_RIS}
	\begin{align}
	&\mathop {\min }\limits_{{\bf{\Theta }}} \left\| {{{\bf{F}}_2}{{\bf{H}}_2}{\bf{\Theta }}{{\bf{H}}_1}{{\bf{F}}_1} - {\bf{W}}} \right\|_F^2 \label{RIS_objectivefunction}\\
	& {\rm{s}}{\rm{.t}}{\rm{. }}~\eqref{RIS_phaseshift}.
	\end{align}
\end{subequations}
Problem \eqref{P1_RIS} is non-convex due to the non-convexity of constraint \eqref{RIS_phaseshift}. To solve it, we first unfold the objective function \eqref{RIS_objectivefunction} as 
\begin{align}
    &\left\| {{{\bf{F}}_2}{{\bf{H}}_2}{\bf{\Theta }}{{\bf{H}}_1}{{\bf{F}}_1} - {\bf{W}}} \right\|_F^2 = \notag\\
    &{\rm{tr}}\left( {{{\bf{F}}_2}{{\bf{H}}_2}{\bf{\Theta }}{{\bf{H}}_1}{{\bf{F}}_1}{\bf{F}}_1^H{\bf{H}}_1^H{{\bf{\Theta }}^H}{\bf{F}}_2^H{\bf{H}}_2^H} \right) + {\rm{tr}}\left( {{\bf{W}}{{\bf{W}}^H}} \right)-\notag\\
 &{\rm{tr}}\left( {{{\bf{F}}_2}{{\bf{H}}_2}{\bf{\Theta }}{{\bf{H}}_1}{{\bf{F}}_1}{{\bf{W}}^H}} \right) - {\rm{tr}}\left( {{\bf{WF}}_1^H{\bf{H}}_1^H{{\bf{\Theta }}^H}{\bf{F}}_2^H{\bf{H}}_2^H} \right). \label{equation_1}
\end{align}
We can rewrite  terms in \eqref{equation_1} in a more compact form given by
\begin{align}
    &{\rm{tr}}\left( {{{\bf{F}}_2}{{\bf{H}}_2}{\bf{\Theta }}{{\bf{H}}_1}{{\bf{F}}_1}{\bf{F}}_1^H{\bf{H}}_1^H{{\bf{\Theta }}^H}{\bf{F}}_2^H{\bf{H}}_2^H} \right)  \notag\\
    &= {\rm{tr}}\left( {{{\bf{\Theta }}^H}{\bf{F}}_2^H{\bf{H}}_2^H{{\bf{F}}_2}{{\bf{H}}_2}{\bf{\Theta }}{{\bf{H}}_1}{{\bf{F}}_1}{\bf{F}}_1^H{\bf{H}}_1^H} \right)\notag\\
    &  = {{\bf{v}}^H}{\bm \Omega} {\bf{v}}, \label{equation_2}
\end{align}
and 
\begin{align}
{\rm{tr}}\left( {{{\bf{F}}_2}{{\bf{H}}_2}{\bf{\Theta }}{{\bf{H}}_1}{{\bf{F}}_1}{{\bf{W}}^H}} \right)& = {\rm{tr}}\left( {{\bf{\Theta }}{{\bf{H}}_1}{{\bf{F}}_1}{{\bf{W}}^H}{{\bf{F}}_2}{{\bf{H}}_2}} \right)\notag\\
& = {{\bf{v}}^T}{\bm{\varphi }}, \label{equation_3}
\end{align}
where  ${\bf{v}} = {\left( {{{\bf{\Theta }}_{1,1}}, \ldots ,{{\bf{\Theta }}_{M,M}}} \right)^T}$, ${\bm \Omega } = \left( {{\bf{F}}_2^H{\bf{H}}_2^H{{\bf{F}}_2}{{\bf{H}}_2}} \right) \odot {\left( {{{\bf{H}}_1}{{\bf{F}}_1}{\bf{F}}_1^H{\bf{H}}_1^H} \right)^T}$,\footnote{Note that 
	${\bm \Omega }$ is a positive semi-definite matrix.} and 
${\bm{\varphi }} = {\left[ {{{\left[ {{{\bf{H}}_1}{{\bf{F}}_1}{{\bf{W}}^H}{{\bf{F}}_2}{{\bf{H}}_2}} \right]}_{1,1}}, \ldots ,{{\left[ {{{\bf{H}}_1}{{\bf{F}}_1}{{\bf{W}}^H}{{\bf{F}}_2}{{\bf{H}}_2}} \right]}_{M,M}}} \right]^T}$.
Based on  \eqref{equation_2} and  \eqref{equation_3}, problem \eqref{P1_RIS} 
can be rewritten as 
\begin{subequations}  \label{P1_RIS_new}
	\begin{align}
	&\mathop {\min }\limits_{\bf{v}} {{\bf{v}}^H}{\bm \Omega} {\bf{v}} - 2{\mathop{\rm Re}\nolimits} \left\{ {{{\bf{v}}^T}{\bm{\varphi }}} \right\} + {\rm{tr}}\left( {{\bf{W}}{{\bf{W}}^H}} \right) \label{RIS_objective_newfunction}\\
	& {\rm{s}}{\rm{.t}}{\rm{. }}~\left| {{{\bf{v}}_i}} \right| = 1,i = 1, \ldots ,M. \label{unit_modulus}
	\end{align}
\end{subequations}
To solve problem \eqref{P1_RIS_new}, a majorization-minimization method can be applied \cite{sone2016sequence}. The key idea  behind the majorization-minimization method is to solve   problem \eqref{P1_RIS_new} by constructing a series of more tractable approximate
objective functions. Specifically,  ${{\bf{v}}^H}{\bm \Omega} {\bf{v}}$ is upper bounded by 
\begin{align}
{{\bf{v}}^H}{\bm \Omega} {\bf{v}} &\le {\lambda _{\max }}M - 2{\mathop{\rm Re}\nolimits} \left\{ {{{\bf{v}}^H}\left( {{\lambda _{\max }}{{\bf{I}}_M} - {\bm \Omega}  } \right){{\bf{v}}^r}} \right\}\notag\\
&+ {{\bf{v}}^{r,H}}\left( {{\lambda _{\max }}{{\bf{I}}_M} - {\bm \Omega}  } \right){{\bf{v}}^r}, \label{phaseshift_upperbound}
\end{align}
where ${{{\bf{v}}^r}}$ represents any initial point of ${{{\bf{v}}}}$ at the $r$-th iteration. Substituting \eqref{phaseshift_upperbound} into \eqref{RIS_objective_newfunction}, problem \eqref{P1_RIS_new} can be  approximated as (ignoring irrelevant constants)
\begin{subequations}  \label{P1_RIS_new_approx}
	\begin{align}
	&\mathop {\min }\limits_{\bf{v}}  - 2{\rm{Re}}\left\{ {{{\bf{v}}^H}\left( {\left( {{\lambda _{\max }}{{\bf{I}}_M} - {\bm \Omega} } \right){{\bf{v}}^r} - {{\bm \varphi} ^*}} \right)} \right\} \\
	& {\rm{s}}{\rm{.t}}{\rm{. }}~\eqref{unit_modulus}.
	\end{align}
\end{subequations}
It is not difficult to prove  that the optimal solution to problem \eqref{P1_RIS_new_approx} is given by 
\begin{align}
	{{\bf{v}}^{{\rm{opt}}}} = {e^{j\arg \left( {\left( {{\lambda _{\max }}{{\bf{I}}_M} - {\bm \Omega} } \right){{\bf{v}}^r} - {{\bm \varphi} ^*}} \right)}}. \label{RISphaseshit_new}
\end{align}

\subsubsection{Overall algorithm}
\begin{algorithm}[!t]
	\caption{Alternating optimization  for solving  problem \eqref{P1}.}	\label{alg1}
	\begin{algorithmic}[1]
		\STATE  \textbf{Initialize} RIS phase-shift vector  ${{\bf{v}}}$ and combiner ${\bf F}_2$.
		\STATE  \textbf{repeat}
		\STATE  \quad Update transmit precoder ${\bf F}_1$ by solving  \eqref{P1_transmitbeamformer}.
		\STATE  \quad Update receive combiner ${\bf F}_2$  based on \eqref{combiner:optimal}
		\STATE \quad Update RIS phase shift  ${{\bf{v}}}$ based on \eqref{RISphaseshit_new}.
		\STATE \textbf{until}  the fractional decrease in the objective value of problem \eqref{P1}  falls below a predefined threshold.
	\end{algorithmic}
\end{algorithm}

We solve each subproblem in an iterative manner where the solution obtained in the current iteration will be the initial point in the next iteration until the convergence is reached. The detailed steps are summarized in Algorithm~\ref{alg1}. 

The computational complexity of Algorithm~\ref{alg1} is calculated as follows: In step $3$, the complexity of calculating the precoder  ${\bf F}_1$ via the Lagrange duality method is ${\cal O}\left( {{{\log }_2}\left( {\frac{{{\lambda ^{{\rm{up}}}}}}{\varepsilon }} \right){N^3}{M^3}} \right)$, where $\varepsilon $ is the predefined accuracy. In  step $4$, a closed-form expression for ${\bf F}_2$ is derived with a  computational complexity of ${\cal O}\left( {{N^3}{M^3}} \right)$.  In  step $5$, a closed-form expression for ${\bf v}$ is derived with a  computational complexity of ${\cal O}\left( {{M^3}} \right)$. Therefore, 
 the overall complexity of Algorithm~\ref{alg1} is given by ${\cal O}\left( {{L_{{\rm{iter}}}}\left( {{{\log }_2}\left( {\frac{{{\lambda ^{{\rm{up}}}}}}{\varepsilon }} \right){N^3}{M^3} + {N^3}{M^3} + {M^3}} \right)} \right)$, where ${{L_{{\rm{iter}}}}}$ denotes the number of iterations required for reaching convergence.

\begin{figure*}[!t]
	\centerline{\includegraphics[width=7.2in]{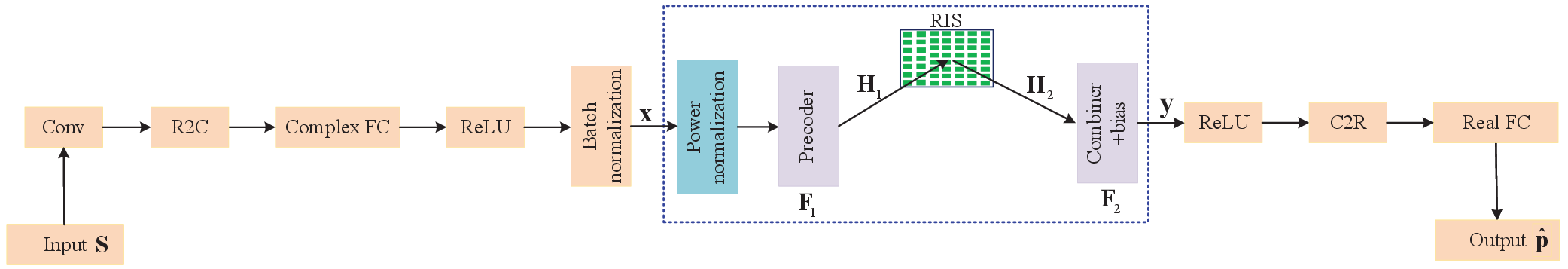}}
	\caption{A novel trainable over-the-air  analog NN architecture.} \label{model_fig2}
\end{figure*}

\section{Trainable RIS-aided Over-the-air  Implementation} \label{section:trainable}
In this section, we instead treat the over-the-air parameters ${\bf F}_1$, ${\bf F}_2$, and $\bf \Theta$ as  trainable, and consider  a novel trainable over-the-air analog NN architecture as depicted in Fig.~\ref{model_fig2}. This architecture includes an input layer, one convolutional (Conv) layer, one R2C layer, one complex FC layer, one C2R layer, one real FC layer, two complex activation layers (ReLU),  one complex batch normalization layer, one power normalization layer, one precoder layer, one RIS layer, one combiner layer, and one output layer. We remark here that our focus in this paper is on the implementation of a FC layer over-the-air using RIS as a computing medium. Extension to more complex network architectures will be considered in future work.  The Conv layer has $2$ output channels with the kernel size of  $3$, stride $4$, and padding $1$;  
 The R2C layer converts the real input value to the complex output value with a dimension equal to half the input dimension, while the C2R layer converts the complex input value to the real output value with  double the input dimension. 
The power normalization layer enforces the transmitter to satisfy the power constraint in \eqref{transmitpower} and   its  output can be described as:
\begin{align}
{{\bf{X}}_{{\rm{out}}}}{\rm{ = }}\frac{{\sqrt {{P_{{\rm{Tx}}}}} }}{{{{\left\| {{{\bf{F}}_1}{\bf{X}}} \right\|}_F}}}{\bf{X}},
\end{align} 
where ${\bf{X}} \in {{\mathbb C}^{N \times B}}$ with $B$ being the batch size and ${{P_{{\rm{Tx}}}}}$ is defined in \eqref{lossfunction}.

In the following, we study two strategies, namely \textit{centralized implementation } and \textit{distributed implementation}, to train the over-the-air analog NN architecture depicted in Fig.~\ref{model_fig2}.

\subsection{Centralized   Training with CSI}
In the centralized approach, we assume that the CSI is available at either the transmitter or the receiver side, and the terminal with the CSI trains   the over-the-air transmission architecture with the loss function defined in \eqref{lossfunction} for the desired task.   Once training is completed, the optimized parameters are then communicated to the other terminal via a dedicated control channel or signaling link. 
In a sense, this approach adapts the network parameters to the current channel, rather than trying to adapt the channel to emulate the target NN as it was done in the previous section.
\subsection{Distributed Over-the-Air  Training without CSI} \label{subsection:distributedtraining}
In the centralized approach, obtaining perfect CSI of channels ${\bf H}_1$ and ${\bf H}_2$ is challenging due to the passive nature of the RIS, which lacks signal processing capabilities for channel estimation. To overcome this limitation, we propose a distributed over-the-air training approach that leverages channel reciprocity and assumes the existence of a reciprocal feedback channel.
 Specifically,  the transmitter transmits in the forward propagation stage, while the receiver transmits over the feedback channel for  gradient propagation in the backward propagation stage. By leveraging channel reciprocity, we propose a low-complexity  implementation, where the gradients of ${\bf F}_1$ and ${\bf F}_2$ are calculated based on  over-the-air computation, reducing  required computational resources at both the transmitter and  receiver. First,  we fix the RIS phase shift matrix to maximize the channel gain (depending on the LoS component), similarly to \cite{hua20233d}. Note that the RIS phase shift matrix only affects the angle of arrival (AoA) from the transmitter to the RIS and the angle of departure (AoD)  from the RIS to the receiver, which can be obtained based on the locations of the transmitter,  RIS, and  receiver;  and hence,  the RIS phase shift matrix does not rely on CSI. Next, we show how to calculate gradients of ${{{\bf{F}}_1}}$ and ${{{\bf{F}}_2}}$ based on  over-the-air computation.

\subsubsection{Over-the-air gradient with respect to (w.r.t.) ${{{\bf{F}}_2}}$}
\begin{table*}[!t]
	\centering
	\caption{ Computation of gradient w.r.t. ${{{\bf{F}}_2}}$}
	\begin{tabular}{|c|c|}
		\hline
		Theoretical gradient              & Over-the-air gradient computation                                                                                               \\ \hline
		\multicolumn{1}{|c|}{$\frac{{\partial {{\cal L}_{{\rm{loss}}}}}}{{\partial {{\bf{F}}_2}}} = \frac{{\partial {{\cal L}_{{\rm{loss}}}}}}{{\partial {\bf{y}}}}{\left( {{{\bf{H}}_2}{\bf{\Theta }}{{\bf{H}}_1}{{\bf{F}}_1}{\bf{x}}{\rm{ + }}{\bf{n}}} \right)^T}$} & \multicolumn{1}{c|}{\begin{tabular}[c]{@{}c@{}}step 1:  transmitter sends signal $\bf x$ to the receiver\\ step 2: receiver receives signal ${{\bf{H}}_2}{\bf{\Theta }}{{\bf{H}}_1}{{\bf{F}}_1}{\bf{x}}{\rm{ + }}{\bf{n}}$\\ step 3: receiver computes $\frac{{\partial {{\cal L}_{{\rm{loss}}}}}}{{\partial {\bf{y}}}}$ and  $\frac{{\partial {{\cal L}_{{\rm{loss}}}}}}{{\partial {\bf{y}}}}{\left( {{{\bf{H}}_2}{\bf{\Theta }}{{\bf{H}}_1}{{\bf{F}}_1}{\bf{x}}{\rm{ + }}{\bf{n}}} \right)^T}$ \end{tabular}} \\ \hline
	\end{tabular} \label{Table: gradientF2}
\end{table*}
According to the chain rule,  the gradient of the loss function w.r.t. ${{{\bf{F}}_2}}$ is given by 
\begin{align}
\frac{{\partial {{\cal  L}_{{\rm{loss}}}}}}{{\partial {{\bf{F}}_2}}} = \frac{{\partial {{\cal L}_{{\rm{loss}}}}}}{{\partial {\bf{y}}}}{\left( {{{\bf{H}}_2}{\bf{\Theta }}{{\bf{H}}_1}{{\bf{F}}_1}{\bf{x}}{\rm{ + }}{\bf{n}}} \right)^T}.  \label{gradient:F2} 
\end{align}
Strictly speaking, $\frac{{\partial {{\cal L}_{{\rm{loss}}}}}}{{\partial {\bf{y}}}}$ can be obtained by a series of chain-rule-based gradient calculations, and we use it for  notation of simplicity.
Computation of $\frac{{\partial {{\cal L}_{{\rm{loss}}}}}}{{\partial {{\bf{F}}_2}}}$ requires a large number of computational resources incurred by  matrix multiplications. To reduce the burden of computational resources, we propose to calculate \eqref{gradient:F2} over-the-air. The detailed steps are described in Table~\ref{Table: gradientF2}: First, the transmitter sends signal $\bf x$ to the receiver. Then, the  receiver receives signal ${{\bf{H}}_2}{\bf{\Theta }}{{\bf{H}}_1}{{\bf{F}}_1}{\bf{x}}{\rm{ + }}{\bf{n}}$. The receiver  computes $\frac{{\partial {{\cal L}_{{\rm{loss}}}}}}{{\partial {\bf{y}}}}{\left( {{{\bf{H}}_2}{\bf{\Theta }}{{\bf{H}}_1}{{\bf{F}}_1}{\bf{x}}{\rm{ + }}{\bf{n}}} \right)^T}$ in \eqref{gradient:F2} .  Note that  ${\frac{{\partial {{\cal L}_{{\rm{loss}}}}}}{{\partial {\bf{y}}}}}$ is available at the receiver since $\bf y$ is known.

\subsubsection{Over-the-air gradient w.r.t. ${{{\bf{F}}_1}}$}
\begin{table*}[!t]
	\centering
	\caption{ Approximate computation of gradient w.r.t. ${{{\bf{F}}_1}}$}
	\begin{tabular}{|c|c|}
		\hline
		Theoretical gradient              & Over-the-air gradient computation                                                                                               \\ \hline
		\multicolumn{1}{|c|}{$\frac{{\partial {{\cal L}_{{\rm{loss}}}}}}{{\partial {{\bf{F}}_1}}} = {\left( {{{\bf{F}}_2}{{\bf{H}}_2}{\bf{\Theta }}{{\bf{H}}_1}} \right)^T}\frac{{\partial {{\cal L}_{{\rm{loss}}}}}}{{\partial {\bf{y}}}}{{\bf{x}}^T}$} & \multicolumn{1}{c|}{\begin{tabular}[c]{@{}c@{}}step 1:  receiver sends signal ${\frac{{\partial {{\cal L}_{{\rm{loss}}}}}}{{\partial {\bf{y}}}}}$ to  transmitter over feedback channel \\ step 2: transmitter receives signal ${{{\left( {{{\bf{F}}_2}{{\bf{H}}_2}{\bf{\Theta }}{{\bf{H}}_1}} \right)}^T}\frac{{\partial {{\cal L}_{{\rm{loss}}}}}}{{\partial {\bf{y}}}} + {\bf{n}}}$\\ step 3: transmitter knows $\bf x$ and performs algebraic computation \\ $\left( {{{\left( {{{\bf{F}}_2}{{\bf{H}}_2}{\bf{\Theta }}{{\bf{H}}_1}} \right)}^T}\frac{{\partial {{\cal L}_{{\rm{loss}}}}}}{{\partial {\bf{y}}}} + {\bf{n}}} \right){{\bf{x}}^T} = {\left( {{{\bf{F}}_2}{{\bf{H}}_2}{\bf{\Theta }}{{\bf{H}}_1}} \right)^T}\frac{{\partial {{\cal L}_{{\rm{loss}}}}}}{{\partial {\bf{y}}}}{{\bf{x}}^T} + {\bf{n}}{{\bf{x}}^T}$ \end{tabular}} \\ \hline
	\end{tabular}\label{Table: gradientF1}
\end{table*} 
According to the chain rule,  the  gradient of the loss function w.r.t. ${{{\bf{F}}_1}}$ is given by 
\begin{align}
\frac{{\partial {{\cal L}_{{\rm{loss}}}}}}{{\partial {{\bf{F}}_1}}} = {\left( {{{\bf{F}}_2}{{\bf{H}}_2}{\bf{\Theta }}{{\bf{H}}_1}} \right)^T}\frac{{\partial {{\cal L}_{{\rm{loss}}}}}}{{\partial {\bf{y}}}}{{\bf{x}}^T}. \label{gradient:F1} 
\end{align}
Approximate computation of \eqref{gradient:F1}  can be achieved  over-the-air  as follows:   First, the receiver  sends signal ${\frac{{\partial {{\cal L}_{{\rm{loss}}}}}}{{\partial {\bf{y}}}}}$ to the transmitter.   
The transmitter receives signal ${{{\left( {{{\bf{F}}_2}{{\bf{H}}_2}{\bf{\Theta }}{{\bf{H}}_1}} \right)}^T}\frac{{\partial {{\cal L}_{{\rm{loss}}}}}}{{\partial {\bf{y}}}} + {\bf{n}}}$, where $\bf n$ is the complex noise component on the feedback channel. Finally, the transmitter  computes  $\left( {{{\left( {{{\bf{F}}_2}{{\bf{H}}_2}{\bf{\Theta }}{{\bf{H}}_1}} \right)}^T}\frac{{\partial {{\cal L}_{{\rm{loss}}}}}}{{\partial {\bf{y}}}} + {\bf{n}}} \right){{\bf{x}}^T} = {\left( {{{\bf{F}}_2}{{\bf{H}}_2}{\bf{\Theta }}{{\bf{H}}_1}} \right)^T}\frac{{\partial {{\cal L}_{{\rm{loss}}}}}}{{\partial {\bf{y}}}}{{\bf{x}}^T} + {\bf{n}}{{\bf{x}}^T}$, since it already has  $\bf x$ . We can observe that this gradient obtained through  over-the-air computation is a noisy version of the desired gradient
 in \eqref{gradient:F1}. The detailed steps are summarized in Table~\ref{Table: gradientF1}.

After receiving the gradient values of ${\bf F}_1$ and  ${\bf F}_2$  computed via over-the-air transmission, each terminal performs standard gradient descent updates on its respective parameters. These updates are iteratively applied without the need for additional coordination.

\section{Extension to  multi-RIS-aided OAC}
\begin{figure*}[!t]
	\centerline{\includegraphics[width=7.2in]{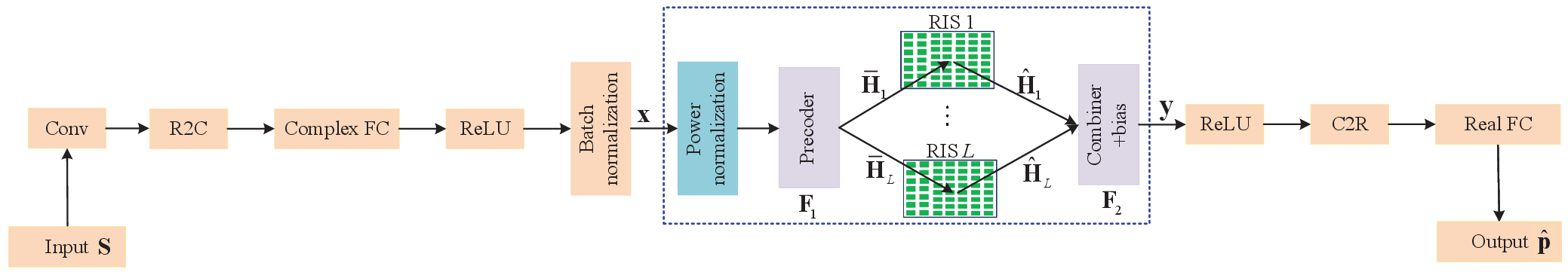}}
	\caption{A multi-RIS aided over-the-air transmission architecture.} \label{model_fig:multiRIS}
\end{figure*}
In the previous sections, we have realized AirFC via a single RIS. However, its  performance is significantly impacted by the channel condition due to the limited degrees of freedom  (DoF) of a single RIS to modify the wireless environment. Specifically,  if $\bf W$ is a high rank  matrix while channels ${\bf H}_1$
and ${\bf H}_2$ are LoS,  we have the following inequalities:
\begin{align}
{\rm{rank}}\left( {{{\bf{F}}_2}{{\bf{H}}_2}{\bf{\Theta }}{{\bf{H}}_1}{{\bf{F}}_1}} \right) &\le \min \left\{ {{\rm{rank}}\left( {{{\bf{H}}_1}} \right),{\rm{rank}}\left( {{{\bf{H}}_2}} \right)} \right\}=1\notag\\
&\ll {\rm{rank}}\left( {\bf{W}} \right),
\end{align}
which indicates that there is a big gap between ${{{\bf{F}}_2}{{\bf{H}}_2}{\bf{\Theta }}{{\bf{H}}_1}{{\bf{F}}_1}}$ and ${\bf{W}}$ no matter how we optimize the  over-the-air parameters  ${\bf F}_1$, ${\bf F}_2$, and $\bf \Theta$.

To solve the rank deficiency problem, we propose to implement AirFC via multiple RISs as shown in Fig.~\ref{model_fig:multiRIS}. In  Fig.~\ref{model_fig:multiRIS}, we assume that there are $L$  RISs separately deployed in the air, each of which  has  $M_i$ 
reflecting elements, $i \in \left\{ {1, \ldots ,L} \right\}$, and satisfy $\sum\limits_{i = 1}^L {{M_i} = M} $. Denote by ${{{\bf{\bar H}}}_i}$ and ${{{\bf{\hat H}}}_i}$ the complex equivalent baseband channel between the transmitter and  RIS $i$ and between the receiver  and  RIS $i$, respectively. In addition, the RIS phase-shift matrix $i$ is denoted by ${{{\bf{\Theta }}_i}}$, $i \in \left\{ {1, \ldots ,L} \right\}$.

The over-the-air channel with multiple RISs is given by (the signal reflected by the RIS more than two times is ignored due to the significant path  attenuation)
\begin{align}
{\bf{H}} = \sum\limits_{i = 1}^L {{{{\bf{\hat H}}}_i}{{\bf{\Theta }}_i}{{{\bf{\bar H}}}_i}} 
 = {\bf{\hat H}}{\rm{diag}}\left( {{{\bf{\Theta }}_1}, \ldots ,{{\bf{\Theta }}_L}} \right){\bf{\bar H}}, \label{channel:multiRIS}
\end{align}
where ${\bf{\hat H}} = \left[ {{{{\bf{\hat H}}}_1}, \ldots ,{{{\bf{\hat H}}}_L}} \right]$ and ${\bf{\bar H}} = {\left[ {{\bf{\bar H}}_1^T, \ldots ,{\bf{\bar H}}_L^T} \right]^T}$.

Comparing  \eqref{channel:multiRIS} with the channel ${{\bf{H}}_2}{\bf{\Theta }}{{\bf{H}}_1}$ in the single-RIS case, we have the following inequalities:
\begin{align}
{\rm{rank}}\left( {\bf{H}} \right) \le \min \left\{ {{\rm{rank}}\left( {\sum\limits_{i = 1}^L {{{{\bf{\hat H}}}_i}} } \right),{\rm{rank}}\left( {\sum\limits_{i = 1}^L {{{{\bf{\bar H}}}_i}} } \right)} \right\}. \label{multiRIS: inequaity1}
\end{align}
In the worst case, where all the channels are LoS, i.e, ${\rm{rank}}\left( {{{{\bf{\hat H}}}_i}} \right) = {\rm{rank}}\left( {{{{\bf{\bar H}}}_i}} \right) = 1$,   $i \in \left\{ {1, \ldots ,L} \right\}$, if $L$ RISs are physically separated, we have 
\begin{align}
{\rm{rank}}\left( {\sum\limits_{i = 1}^L {{{{\bf{\hat H}}}_i}} } \right) = L,~~ {{\rm{rank}}\left( {\sum\limits_{i = 1}^L {{{{\bf{\bar H}}}_i}} } \right)}=L.
\end{align}
 Then, \eqref{multiRIS: inequaity1} can be further deduced as 
 \begin{align}
{\rm{rank}}\left( {\bf{H}} \right) \le L,
 \end{align}
 which indicates that if $L$ is large, the rank of channel $\bf H$ can  approach  that  of $\bf W$. Therefore, by jointly optimizing ${\bf F}_1$, ${\bf F}_2$, and $\bf \Theta$, the emulation error is expected to be reduced significantly.

\textbf{\textit{ Remark:}} Channel expressions for multi-RIS and single-RIS cases are similar,  proposed  algorithms/implementations for the single-RIS case can be easily extended to the multi-RIS case. Thus, the details of the proposed algorithm/implementation for the multi-RIS case is omitted here for brevity.
 


\section{Numerical Results}
In this section, we provide numerical results to evaluate the performance of the proposed AirFC scheme aided by RIS.  We consider  MNIST  and Fashion MNIST datasets as proof of concept. Both datasets contain $10$ classes, and each image  is of dimensions  $28\times28$.  The number of epochs and the batch size are set to $200$ and $32$, respectively.  Unless otherwise stated, we set  $N=49$,
${\lambda _{\rm{p}}}={\lambda _{\rm{RIS}}}=100$,  and  ${\sigma ^2} = 1$. A widely adopted Rician fading channel model with Rician factor $K$ is considered.

\subsection{Single RIS}
We first study the emulation error using a single RIS, and then study the classification accuracy performance of AirFC based on the MNIST dataset.
\subsubsection{Over-the-air parameters computation} We first train a target  NN  with an architecture as shown in Fig.~\ref{model_fig1}, and obtain the well trained middle layer weights $\bf W$. Then, for each channel realization, we compute the over-the-air parameters, namely, ${\bf F}_1$, ${\bf F}_2$, and $\bf \Theta$,  according to Algorithm~\ref{alg1}. To show the impact of transmit power $P_{\rm max}$, number of RIS reflecting elements $M$, and Rician factor $K$ on the system performance, the following schemes are considered:
  \begin{itemize}
	\item Sum error, $\left| {{{\bf{v}}_i}} \right| = 1$ $\left( {\left| {{{\bf{v}}_i}} \right| \le 1} \right)$: This represents the sum error comprising both weight and bias components, calculated according to equation \eqref{objectivefunction}. When $\left| {{{\bf{v}}_i}} \right| = 1$, the amplitude of the RIS phase shift is fixed; when $\left| {{{\bf{v}}_i}} \right| \le 1$, both amplitude and phase can be jointly optimized. 
	\item Weight error, $\left| {{{\bf{v}}_i}} \right| = 1$ $\left( {\left| {{{\bf{v}}_i}} \right| \le 1} \right)$: This quantifies the weight mismatch and is calculated based on  $\left\| {{{\bf{F}}_2}{{\bf{H}}_2}{\bf{\Theta }}{{\bf{H}}_1}{{\bf{F}}_1} - {\bf{W}}} \right\|_F^2$ according to \eqref{objectivefunction}.
	\item Bias error, $\left| {{{\bf{v}}_i}} \right| = 1$ $\left( {\left| {{{\bf{v}}_i}} \right| \le 1} \right)$: This represents the bias contribution, calculated by ${{\mathbb E}_{\bf{n}}}\left\{ {{{\left\| {{{\bf{F}}_2}{\bf{n}}} \right\|}^2}} \right\}$ according to \eqref{objectivefunction}.
\end{itemize}
 \begin{figure}[!t]
 	\centerline{\includegraphics[width=3.2in]{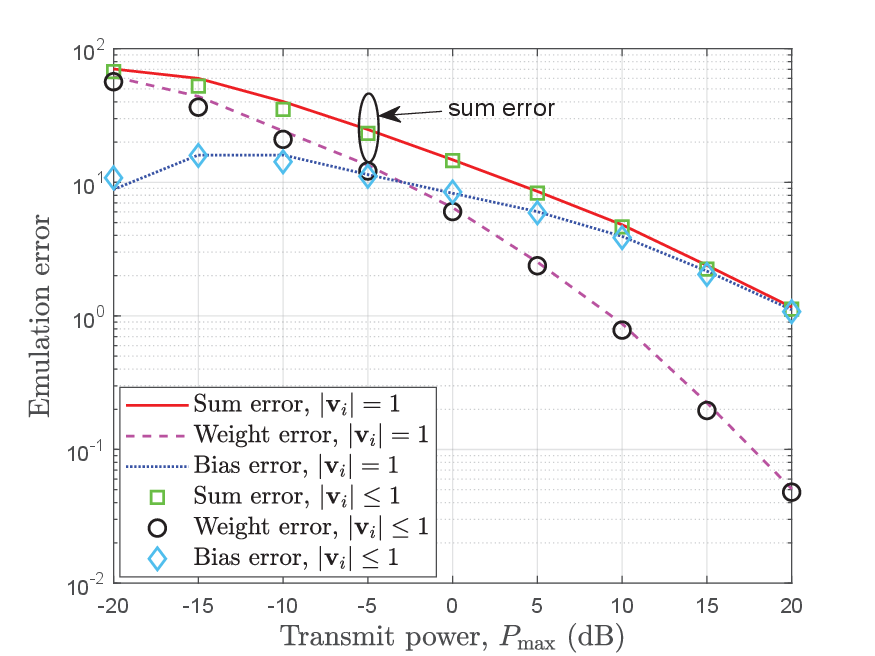}}
 	\caption{Transmit power $P_{\rm max}$  versus  emulation error under $K=10~{\rm dB}$ and $M=100$.} \label{SingleRISOAcal:fig1}
 	\vspace{-0.4cm}
 \end{figure}

In Fig.~\ref{SingleRISOAcal:fig1}, we study the impact of the transmit power $P_{\rm max}$ on the emulation error for $K=10~{\rm dB}$  and $M=100$.  It is observed that the sum emulation error significantly decreases as $P_{\rm max}$ increases.  This is expected because a higher $P_{\rm max}$ provides more DoF for the optimization of ${\bf F}_1$. In addition, it can be seen that there exists a small performance gap between $\left| {{{\bf{v}}_i}} \right| = 1$ and $ {\left| {{{\bf{v}}_i}} \right| \le 1}$  in the ``Sum error'' scheme when $P_{\rm max}$ is small. For example, under $P_{\rm max}=-20~{\rm dB}$, the sum error for $\left| {{{\bf{v}}_i}} \right| = 1$ is about $70.5$,  while that for $ {\left| {{{\bf{v}}_i}} \right| \le 1}$  is about $67.4$, indicating that the optimal amplitude of RIS elements is not necessarily equal to $1$ under low $P_{\rm max}$.
 As $P_{\rm max}$ increases, the performance gap between $\left| {{{\bf{v}}_i}} \right| = 1$ and $ {\left| {{{\bf{v}}_i}} \right| \le 1}$ gradually vanishes. This suggests that the amplitudes of RIS elements tend toward $1$ as $P_{\rm max}$ increases. Moreover,  both the weight  and bias errors significantly decrease with increasing  $P_{\rm max}$,  implying that higher $P_{\rm max}$ facilitates better emulation of the digital FC layer.

 \begin{figure}[!t]
	\centerline{\includegraphics[width=3.2in]{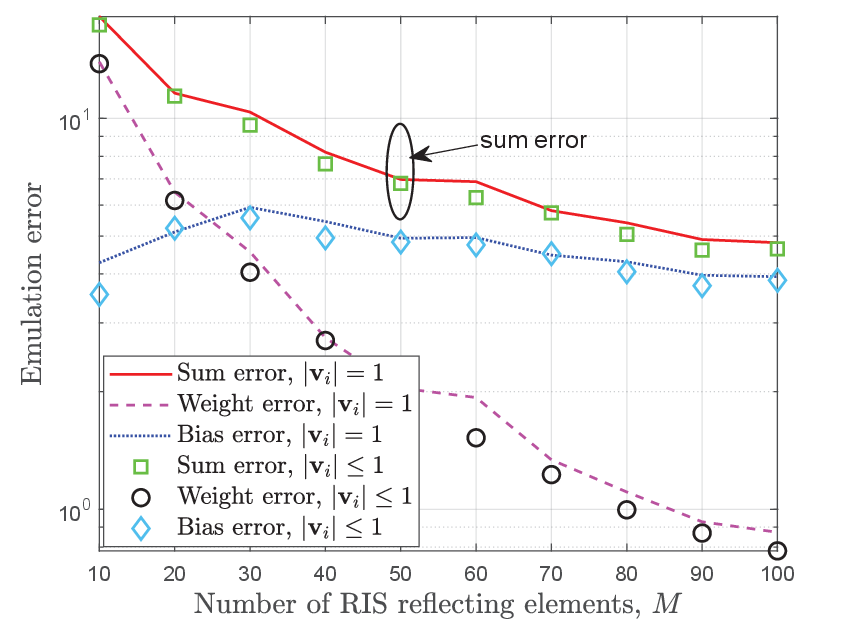}}
	\caption{Number of RIS reflecting elements $M$  versus  emulation error under $K=10~{\rm dB}$ and $P_{\rm max} =10~{\rm dB}$.} \label{SingleRISOAcal:fig2}
		\vspace{-0.4cm}
\end{figure}

In Fig.~\ref{SingleRISOAcal:fig2}, we study the impact of the number of RIS reflecting elements $M$  on the emulation error under $K=10~{\rm dB}$ and $P_{\rm max} =10~{\rm dB}$. It is observed that as $M$ increases, the emulation error decreases. This is because more reflecting elements contribute to higher passive beamforming gains, thereby providing additional DoF for the joint optimization of ${\bf F}_1$ and ${\bf F}_2$, which leads to reduced emulation error. In addition, the scheme with  $ {\left| {{{\bf{v}}_i}} \right| \le 1}$   slightly  outperforms that with  $ {\left| {{{\bf{v}}_i}} \right| = 1}$. It is also observed that the bias error does not decrease monotonically with $M$.
This can be attributed to the fact that the optimization process minimizes the combined sum of weight and bias errors, rather than the bias error alone. Consequently, when 
$M$ is small, the weight error tends to dominate the overall emulation error.

 \begin{figure}[!t]
	\centerline{\includegraphics[width=3.2in]{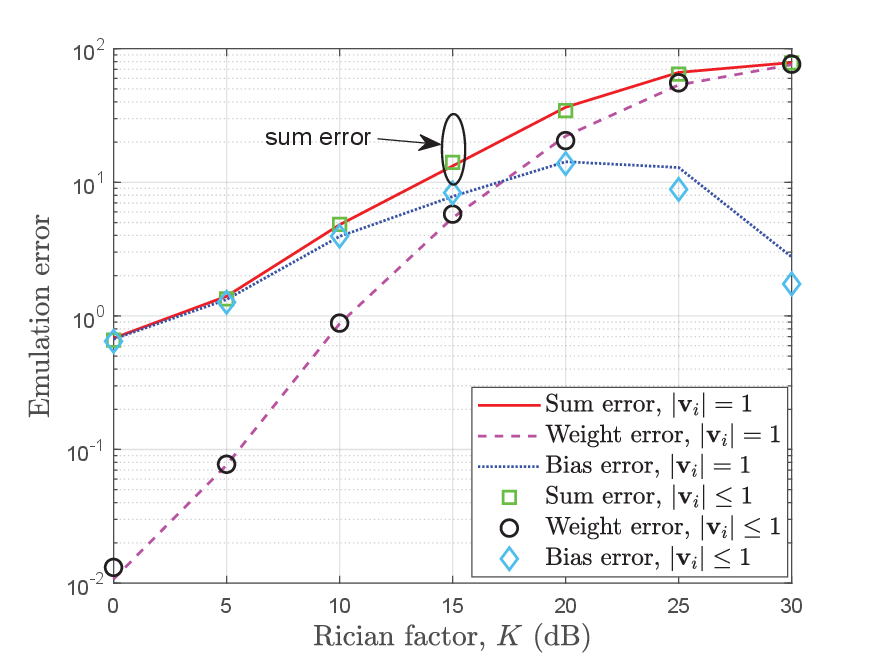}}
	\caption{ Emulation error versus Rician factor $K$ under $P_{\rm max} =10~{\rm dB}$ and $M=100$.} \label{SingleRISOAcal:fig3}
		\vspace{-0.4cm}
\end{figure}

Fig.~\ref{SingleRISOAcal:fig3} plots the emulation error against the Rician factor $K$.  As $K$ is small, the ranks of channels ${\bf H}_1$ and ${\bf H}_2$  are high. However, with the increase in $K$, the LoS path is dominated and the ranks of channels ${\bf H}_1$ and ${\bf H}_2$ decay significantly, which implies that the rank of ${{{\bf{F}}_2}{{\bf{H}}_2}{\bf{\Theta }}{{\bf{H}}_1}{{\bf{F}}_1}}$ is always smaller than that of ${\bf{W}}$ no matter how we optimize ${{{\bf{F}}_1}}$, ${{{\bf{F}}_2}}$, and ${\bf{\Theta }}$,  inevitably causing large emulation error. In addition, one can observe that the bias error drops when $K\ge 20~{\rm dB}$, which indicates that the weight error  dominates the bias error when $K\ge 20~{\rm dB}$.
\subsubsection{Classification accuracy}
To evaluate the impact of  parameters ${\bf F}_1$, ${\bf F}_2$, and $\bf \Theta$  on the classification accuracy, we consider the inference of samples from the MNIST dataset. For each batch, a channel sample is drawn, and the corresponding parameters are computed using  Algorithm~\ref{alg1}. We consider two RIS phase shift models: $\left| {{{\bf{v}}_i}} \right| = 1$ and  $ {\left| {{{\bf{v}}_i}} \right| \le 1}$.

\begin{figure}[!t]
	\centerline{\includegraphics[width=3.2in]{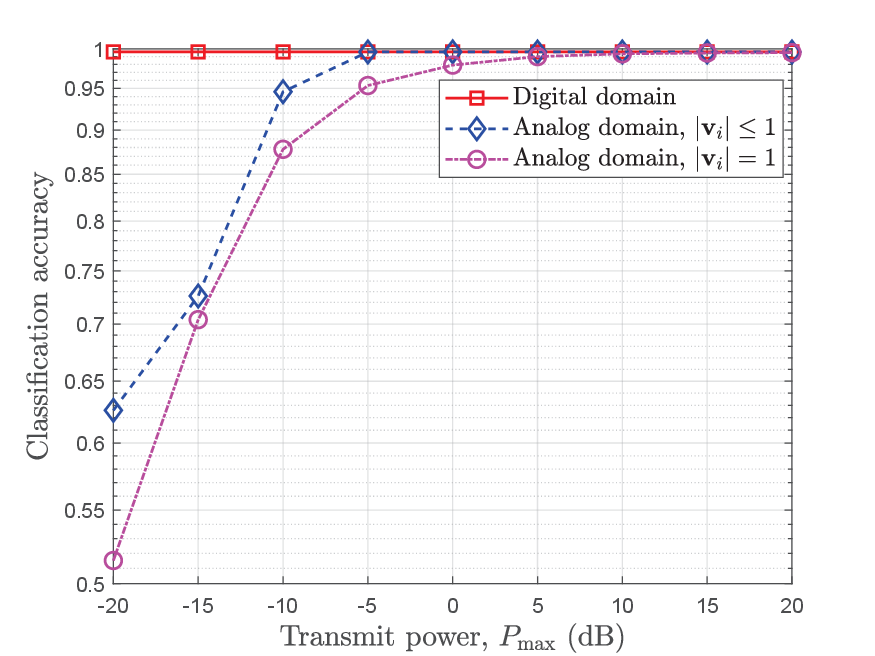}}
	\caption{  Classification accuracy versus $P_{\rm max}$ under $K=10~{\rm dB}$ and $M=100$.} \label{SingleRISOAcal:fig4}
		\vspace{-0.4cm}
\end{figure}

Fig.~\ref{SingleRISOAcal:fig4} plots the classification accuracy  as a function of $P_{\rm max}$ under $K=10~{\rm dB}$ and $M=100$. 
The digital scheme  shows the accuracy of the target NN, which serves as an upper bound on the performance achieved by the over-the-air implementation.
It can be seen that the performance gap between the upper bound and the analog implementation is significant when $P_{\rm max}$ is small, but it gradually diminishes as   $P_{\rm max}$ increases. 
This behavior is attributed to the fact that when  $P_{\rm max}$ is small, the introduced  bias noise power ${\left\| {{{\bf{F}}_2}{\bf{n}}} \right\|^2}$  is relatively large, leading to poor performance. As $P_{\rm max}$ increases,  ${\left\| {{{\bf{F}}_2}} \right\|_F^2}$ can be effectively reduced to better match the target  weights, i.e., minimizing $\left\| {{{\bf{F}}_2}{{\bf{H}}_2}{\bf{\Theta }}{{\bf{H}}_1}{{\bf{F}}_1} - {\bf{W}}} \right\|_F^2$,  while simultaneously reducing the bias noise ${\left\| {{{\bf{F}}_2}{\bf{n}}} \right\|^2}$. Moreover, we can observe that when $P_{\rm max} \le 0 ~{\rm dB}$, the classification accuracy achieved by ${\left| {{{\bf{v}}_i}} \right| \le 1} $ is slightly higher than that achieved by ${\left| {{{\bf{v}}_i}} \right| = 1} $, which is consistent with the results shown in  Fig.~\ref{SingleRISOAcal:fig1}.

 \begin{figure}[!t]
	\centerline{\includegraphics[width=3.2in]{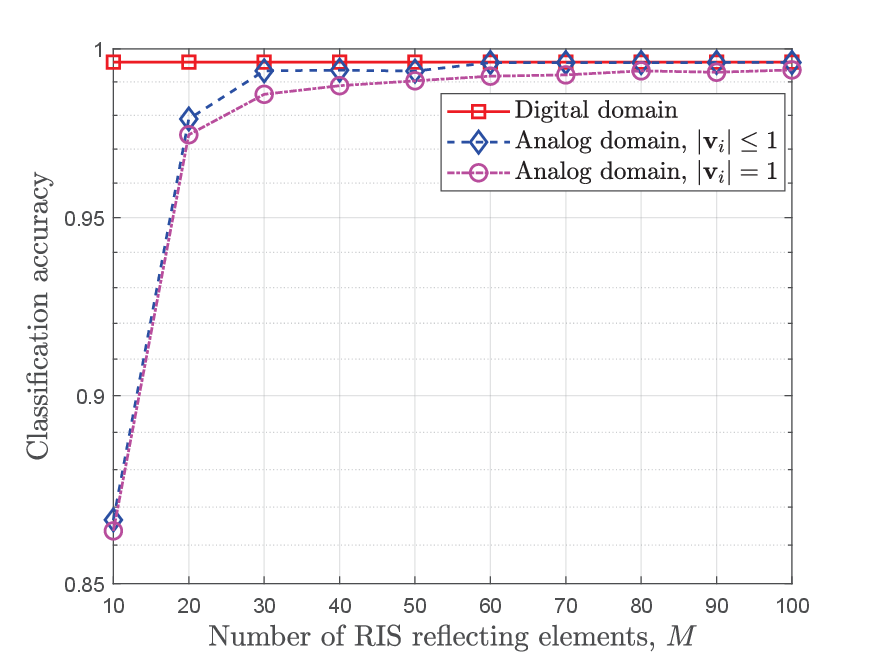}}
	\caption{   Classification accuracy versus $M$  under $K=10~{\rm dB}$ and $P_{\rm max} =10~{\rm dB}$.} \label{SingleRISOAcal:fig5}
		\vspace{-0.4cm}
\end{figure}

Fig.~\ref{SingleRISOAcal:fig5} depicts  the  classification accuracy  versus $M$  under $K=10~{\rm dB}$ and $P_{\rm max} =10~{\rm dB}$. It is observed that when $M$ is small, e.g.,  $M\le20$, the classification error obtained by the analog  scheme is low. This is due to two reasons. First, a small $M$ indicates that the rank of ${{{\bf{F}}_2}{{\bf{H}}_2}{\bf{\Theta }}{{\bf{H}}_1}{{\bf{F}}_1}}$ is always small. Second, a small $M$  indicates that the passive beamforming gain provided by the RIS is small, which means that the introduced bias error ${\left\| {{{\bf{F}}_2}{\bf{n}}} \right\|^2}$  would be large. As $M$ increases, the rank of the effective channel also increases, and a higher passive beamforming gain is provided against noise, resulting in high classification accuracy. 

\subsubsection{Trainable over-the-air parameters}
Next, we consider training parameters ${\bf F}_1$, ${\bf F}_2$, and $\bf \Theta$ for each channel realization for the over-the-air transmission architecture given in Fig.~\ref{model_fig2}. To show the performance gain obtained by the proposed scheme, the following schemes are compared:\footnote{The benchmark where the RIS phase shifts are randomly set is not considered, as the resulting poor classification performance is straightforward.}
\begin{itemize}
    \item Trainable, ${\left| {{{\bf{v}}_i}} \right| \le 1}$: In this scheme, we consider the ideal RIS reflection model, and the parameters are trained based on the loss function given by ${{\cal L}_{{\rm{loss}}}} =  - \sum\limits_{i = 1}^C {{p_i}\log \left( {{{\hat p}_i}} \right)}  - {\lambda _{\rm{p}}}\min \left\{ {0,{P_{\max }} - {P_{{\rm{Tx}}}}} \right\} - {\lambda _{{\rm{RIS}}}}\sum\limits_{i = 1}^M {\min \left\{ {0,1 - \left| {{{\bf v}_i}} \right|} \right\}} $.
    \item Trainable, $\left| {{{\bf{v}}_i}} \right| = 1$: In this scheme, we consider the practical RIS reflection model, and the parameters are trained based on the loss function given by \eqref{lossfunction}.
    \item Baseline $1$: In this scheme, the RIS phase shift matrix is non-trainable and is fixed to maximize the channel gain, while  ${\bf F}_1$  and  ${\bf F}_2$ are trained based on the loss function given by \eqref{lossfunction}.
    \item Baseline $2$: This schemes corresponds to the distributed training scheme  in  Section~\ref{subsection:distributedtraining}.
\end{itemize}

\begin{figure}[!t]
	\centerline{\includegraphics[width=3.2in]{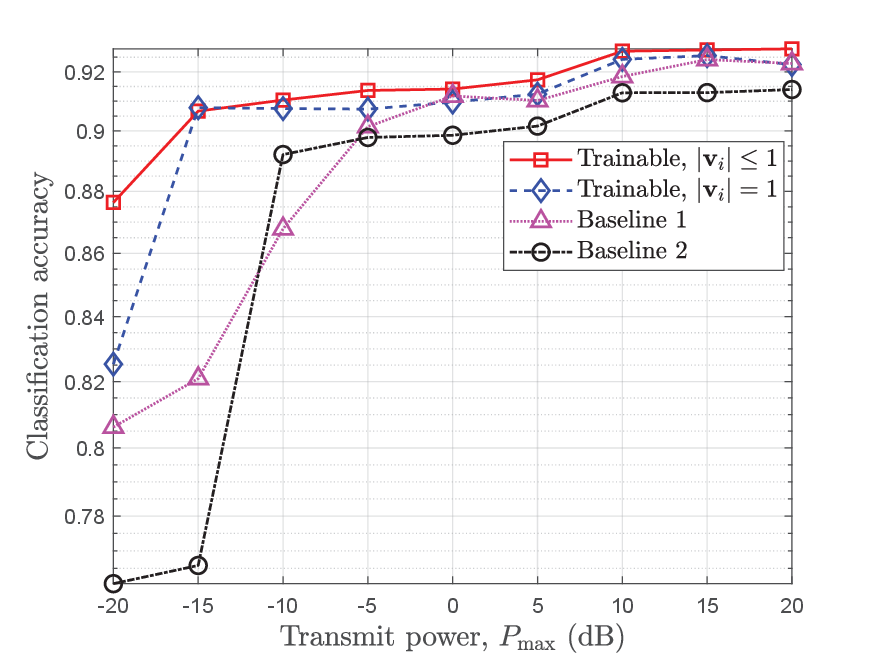}}
	\caption{   Classification accuracy versus $P_{\rm max}$  under $K=-10~{\rm dB}$ and $M=50$.} \label{SingleRISOAcal:fig6}
		\vspace{-0.4cm}
\end{figure}

In Fig.~\ref{SingleRISOAcal:fig6}, we study the impact of $P_{\rm max}$ on the classification accuracy  for different schemes under $K=-10~{\rm dB}$ and $M=50$. It can be seen that ``Trainable, ${\left| {{{\bf{v}}_i}} \right| \le 1}$'' scheme achieves the best classification accuracy. It can also be observed that when $P_{\rm max} < -5~{\rm dB}$, the 
classification accuracy gap between ``Trainable, ${\left| {{{\bf{v}}_i}} \right| = 1}$'' scheme and Baseline $1$ is large, while the gap diminishes as $P_{\rm max}$ increases. This indicates that the RIS phase shift obtained by ``Trainable, ${\left| {{{\bf{v}}_i}} \right| = 1}$''  also maximizes the effective channel gain. Furthermore, when $P_{\rm max}$ is small,  Baseline $2$  achieves the poorest performance  due to the relatively strong noise incurred at the calculation of the gradient w.r.t. ${\bf F}_1$,  which hinders training of the parameters. With the increase in $P_{\rm max}$, the impact of the noise incurred in gradient calculation  vanishes, and the classification accuracy is significantly improved.

\begin{figure}[!t]
	\centerline{\includegraphics[width=3.2in]{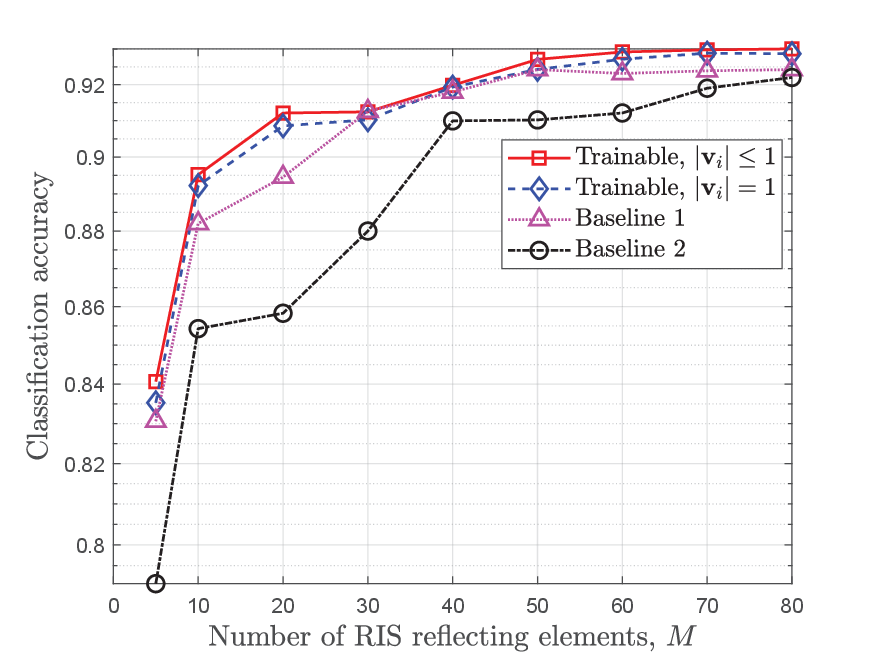}}
	\caption{Classification accuracy versus $M$ under $K=-10~{\rm dB}$ and $P_{\rm max}=10~{\rm dB}$.} \label{SingleRISOAcal:fig7}
		\vspace{-0.4cm}
\end{figure}

In Fig.~\ref{SingleRISOAcal:fig7}, we study the classification accuracy obtained by the over-the-air transmission architecture  versus $M$ under $K=-10~{\rm dB}$ and $P_{\rm max}=10~{\rm dB}$. It is observed that the classification accuracy obtained by all the schemes increases with  $M$. This is because more reflecting elements provide higher passive beamforming gain against the noise. In addition, more reflecting elements can also weaken the impact of noise on the gradient computation in Baseline $2$.

\begin{figure}[!t]
	\centerline{\includegraphics[width=3.2in]{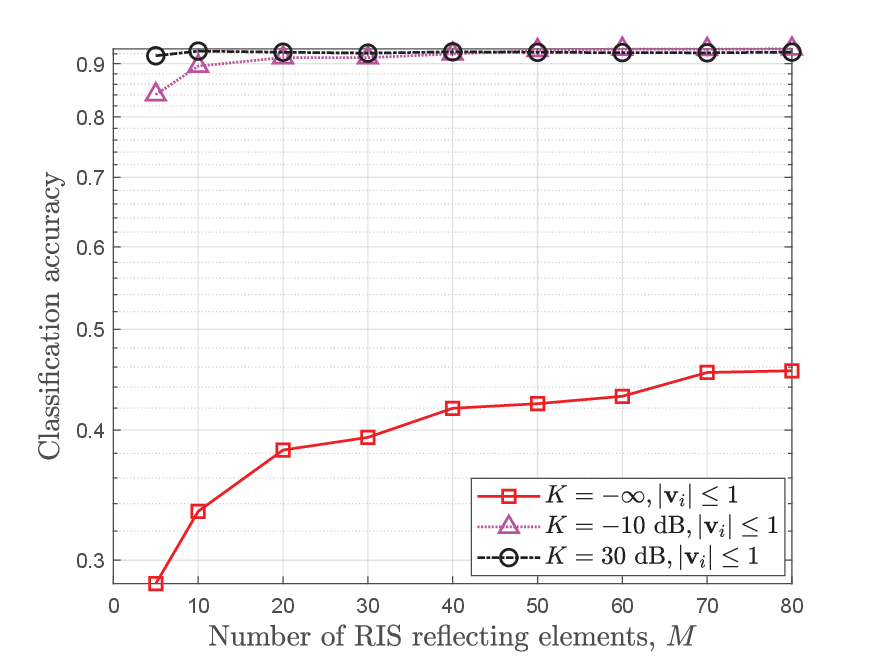}}
	\caption{Classification accuracy versus $M$ for different $K$ values for  $P_{\rm max}=10~{\rm dB}$.} \label{SingleRISOAcal:fig8}
		\vspace{-0.4cm}
\end{figure}

In Fig.~\ref{SingleRISOAcal:fig8}, we study the classification accuracy versus $M$ for different $K$ values for  $P_{\rm max}=10~{\rm dB}$. Here, $K =  - \infty $ corresponds to the Rayleigh fading channel. It is observed that for 
$K =  - \infty $,  classification accuracy remains low even when $M$ is large. For example, when $M=80$,  classification accuracy is approximately $45$\%. This is because the Rayleigh fading channel is non-directional in space, weakening the beamforming effect. In contrast, for larger $K$,  the channel experiences LoS, allowing more energy to be concentrated along a dominant path by designing ${\bf F}_1$, ${\bf F}_2$, and $\bf \Theta$. However, this observation differs from that in Fig.~\ref{SingleRISOAcal:fig3}. Specifically, the parameters in Fig.~\ref{SingleRISOAcal:fig3} are non-trainable and computed using Algorithm~\ref{alg1}, while those in Fig.~\ref{SingleRISOAcal:fig8} are obtained through training. This discrepancy indicates that, in the training case,  parameters are influenced not only by the middle FC layer but also by other NN components.
From  Fig.~\ref{SingleRISOAcal:fig8}, it can be inferred that when the channel power gain is large (i.e., $K$ is large),  the impact of noise becomes negligible.

%
%

\subsection{Multiple RISs}
In this subsection, we investigate the multi-RIS scenario based on the MNIST dataset and explore the effect of the potential spatial diversity gain on  classification accuracy. To this end, we first examine the performance of multi-RIS in emulating the given FC layer, followed by an evaluation of classification accuracy through  joint training of  ${\bf F}_1$, ${\bf F}_2$, and ${\bf \Theta}_i$, $i = 1, \ldots ,L$. In the following analysis, we consider practical RIS phase shifts where the amplitude of each RIS reflecting element is set to $1$.
\subsubsection{Multi-RIS over-the-air computation}

 \begin{figure}[!t]
	\centerline{\includegraphics[width=3.2in]{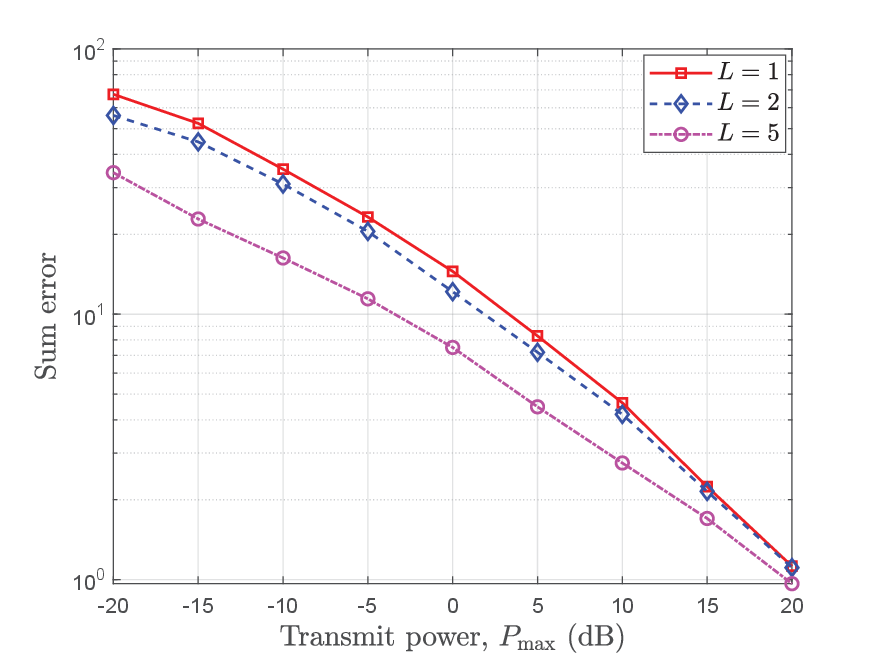}}
	\caption{   Sum emulation error versus  $P_{\rm max}$ for different $L$ values under $K=10~{\rm dB}$ and $M=100$.} \label{SingleRISOAcal:fig10}
		\vspace{-0.4cm}
\end{figure}

In Fig.~\ref{SingleRISOAcal:fig10}, we study the relationship between $P_{\rm max}$  and the sum emulation error for  $L=1, 2$, and $5$.  It is observed that a larger  $L$ leads to a lower emulation error, particularly when $P_{\rm max}$ is small.   For example, at $P_{\rm max} = -20~{\rm dB}$, the sum error for $L=5$ is $34.1$, compared to   $67.4$ for  $L=1$, reflecting a reduction of approximately $49.5\%$. This improvement stems from the spatial diversity gain achieved by independently deployed RISs, which increases the effective channel rank.

 \begin{figure}[!t]
	\centerline{\includegraphics[width=3.2in]{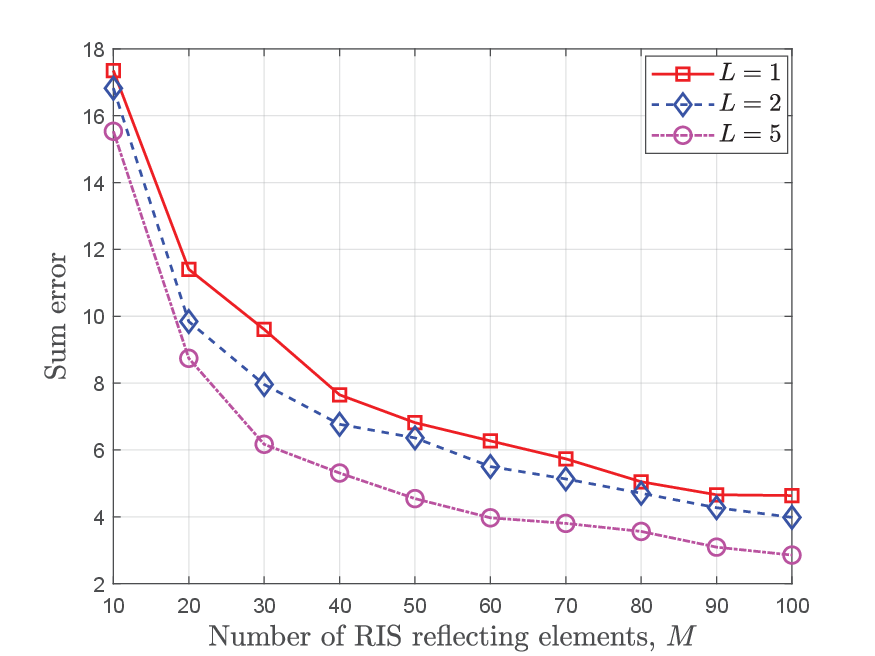}}
	\caption{Sum emulation error  versus $M$   for $L$ under $K=10~{\rm dB}$ and $P_{\rm max} = 10~{\rm dB}$.}  \label{SingleRISOAcal:fig11}
		\vspace{-0.4cm}
\end{figure}

Fig.~\ref{SingleRISOAcal:fig11} investigates the sum error versus $M$ for different number of RISs $L=1, 2$, and $5$. It is evident that a larger  $M$ results in a lower emulation error due to the increased passive gain provided by the RISs. In addition, a larger 
$L$ consistently yields a lower emulation error, especially when  $M$ is large. For example, when  $M=100$, the sum error for $L=5$ is $2.8$, while that for $L=1$ is about $4.6$,  indicating an error reduction of approximately  $39.1\%$.

 \begin{figure}[!t]
	\centerline{\includegraphics[width=3.2in]{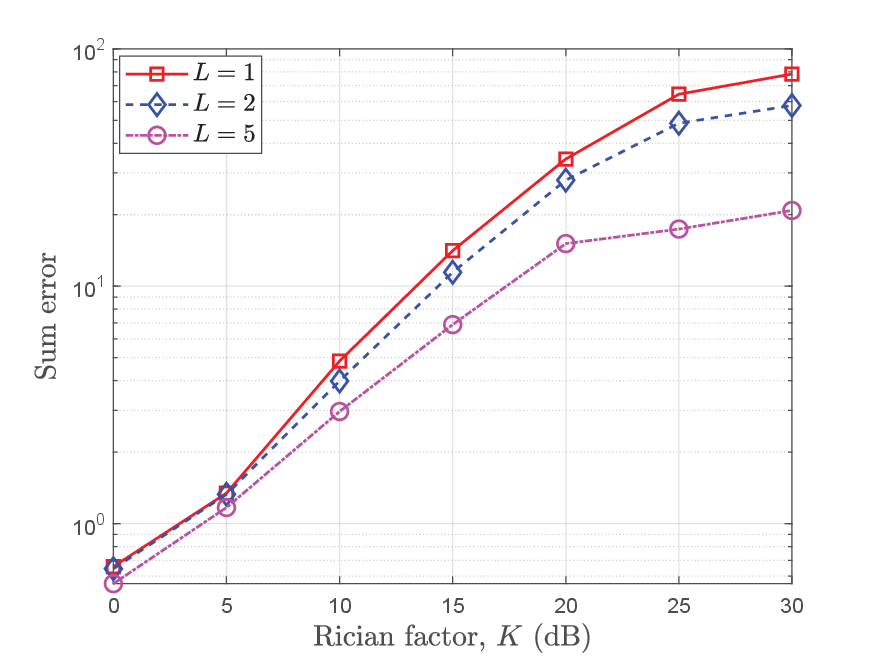}}
	\caption{Sum emulation error versus $K$ for different $L$ values under $M=100$ and $P_{\rm max} = 10~{\rm dB}$.}  \label{SingleRISOAcal:fig12}
		\vspace{-0.4cm}
\end{figure}

Fig.~\ref{SingleRISOAcal:fig12} investigates the impact of $K$  on the sum emulation error for different values of  $L$. As $K$  becomes large, the channel tends to be dominated by the LoS component, and the NLoS components can be neglected, leading to a deficiency in channel rank. Thus, the emulation error increases significantly with $K$. However, by deploying multiple RISs, additional LoS paths are introduced, effectively mitigating the rank deficiency and reducing the emulation error.  For example, when  $K = 30~{\rm dB}$, the sum error for $L=5$ is $20.8$, compared to $78.3$ for $L=1$, corresponding to an emulation error reduction of approximately $73.4\%$.

\begin{figure}[!t]
	\centerline{\includegraphics[width=3.2in]{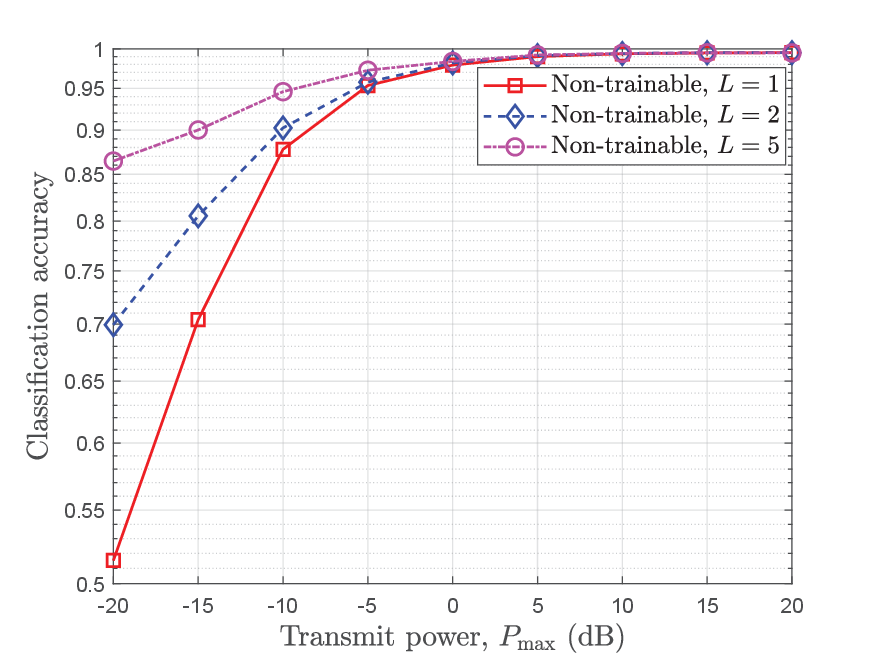}}
	\caption{ Classification accuracy versus $P_{\rm max}$   for different $L$ values under $K=10~{\rm dB}$ and $M=100$.} \label{SingleRISOAcal:fig13}
		\vspace{-0.4cm}
\end{figure}

 \begin{figure}[!t]
	\centerline{\includegraphics[width=3.2in]{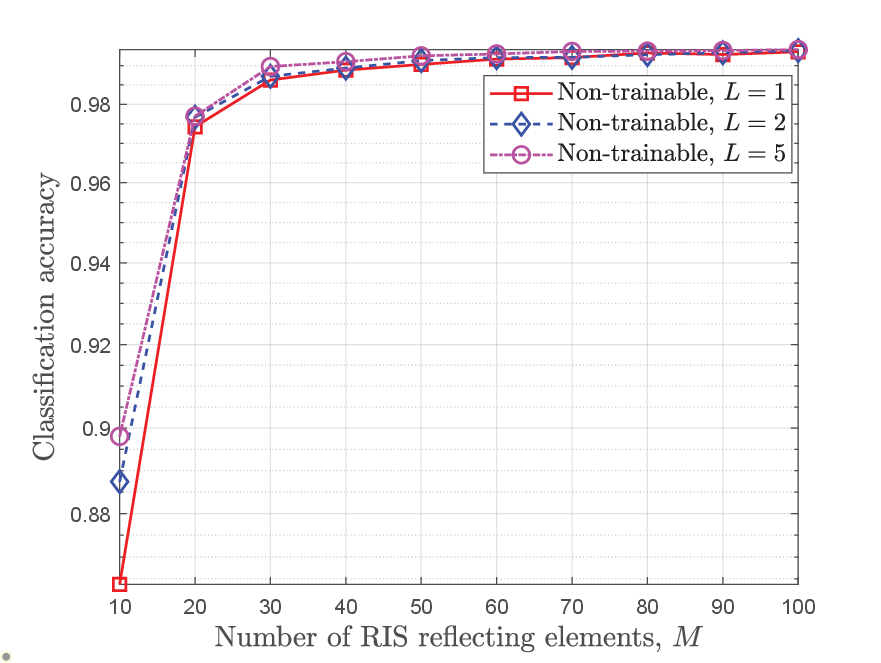}}
	\caption{Classification accuracy  versus $M$  for different $L$ values under $K=10~{\rm dB}$ and $P_{\rm max} = 10~{\rm dB}$.}  \label{SingleRISOAcal:fig14}
		\vspace{-0.4cm}
\end{figure}

 \begin{figure}[!t]
	\centerline{\includegraphics[width=3.2in]{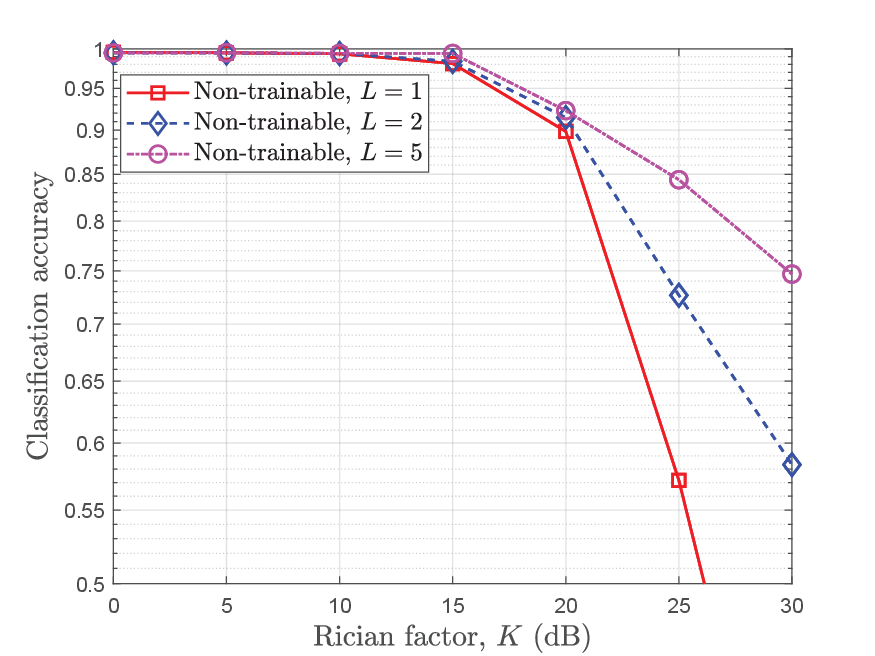}}
	\caption{Classification accuracy  versus $K$   for different $L$  values under $M=100$ and $P_{\rm max} = 10~{\rm dB}$.}  \label{SingleRISOAcal:fig15}
		\vspace{-0.4cm}
\end{figure}

Then, we replace the target FC layer with the over-the-air transmission layer. The impacts of  $P_{\rm max} $,  $M$, and $K$ on the classification performance for different $L$ values
are studied in Fig.~\ref{SingleRISOAcal:fig13}, Fig.~\ref{SingleRISOAcal:fig14}, and Fig.~\ref{SingleRISOAcal:fig15}, respectively.   It can be observed from these figures that a larger  $L$ consistently leads to improved classification performance. This is because a larger $L$ achieves a lower emulation error, as demonstrated previously in  Fig.~\ref{SingleRISOAcal:fig10}, Fig.~\ref{SingleRISOAcal:fig11}, and Fig.~\ref{SingleRISOAcal:fig12}, respectively.
\subsubsection{Multi-RIS over-the-air training}
Next, we investigate the case where all over-the-air parameters are trainable.
 \begin{figure}[!t]
	\centerline{\includegraphics[width=3.2in]{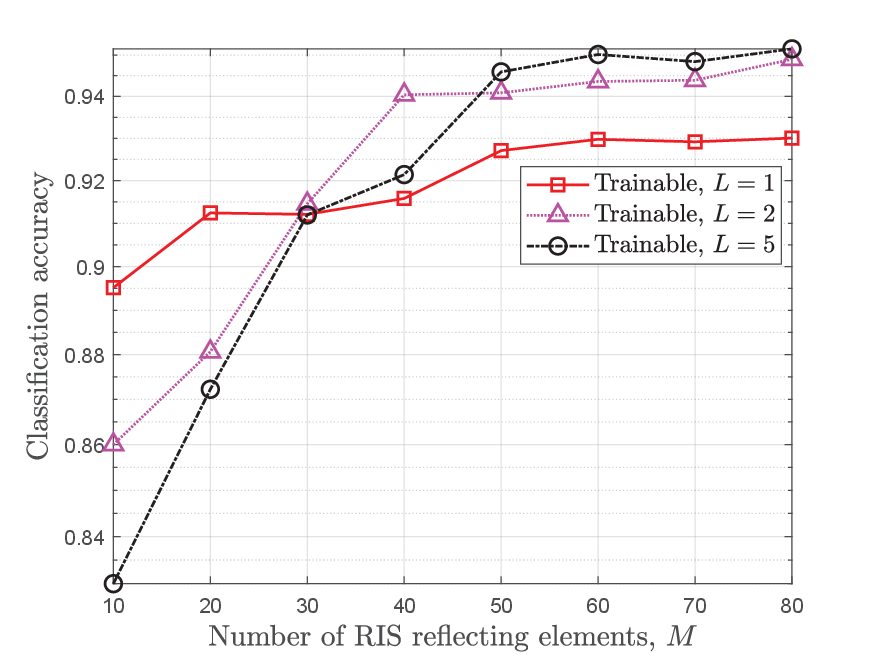}}
	\caption{ Classification accuracy  versus $M$     for different $L$ values under $P_{\rm max} = 10~{\rm dB}$ and $K = 10~{\rm dB}$.}  \label{SingleRISOAcal:fig16}
		\vspace{-0.4cm}
\end{figure}
In Fig.~\ref{SingleRISOAcal:fig16}, we study  $M$   versus  classification accuracy for different $L$. It is observed that the classification accuracy increases monotonically with $M$ for all the schemes, which is expected since a larger $M$ provides higher passive gain against noise. Interestingly, when $M$ is small, e.g., $M\le 30$, the classification accuracy achieved with $L=5$ is lower than that with $L=1$ and $L=2$. However, when $M \ge 50$, the classification accuracy obtained by $L=5$ performs best.  This phenomenon can be explained as follows: for a given total number of reflecting elements, a larger $L$ implies that the number of reflecting elements per RIS is 
$M/L$. When $M$ is small, each RIS has fewer reflecting elements, resulting in limited passive beamforming gain and thus degraded performance. Moreover, the spatial-diversity gain is negligible in this case. In contrast, as $M$ increases, the spatial-diversity gain provided by multiple RISs becomes more prominent, significantly enhancing the effective channel gain against noise. As a result, when 
$M$ is sufficiently large, the classification performance for 
$L=5$ surpasses that for $L=1$ and $L=2$.

 \begin{figure}[!t]
	\centerline{\includegraphics[width=3.2in]{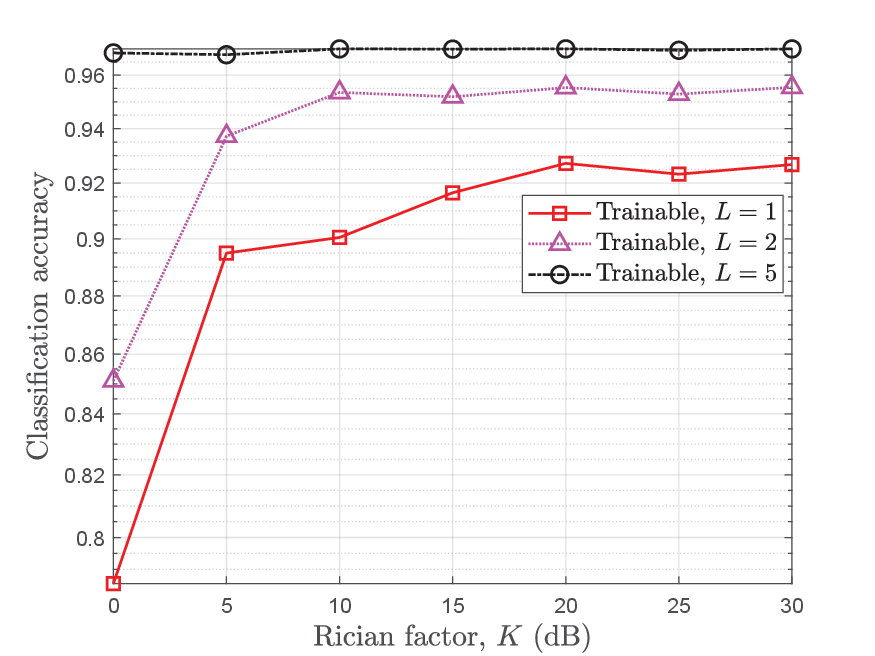}}
	\caption{Classification accuracy  versus $K$ for different $L$ values under $M=100$ and $P_{\rm max} = 10~{\rm dB}$.}  \label{SingleRISOAcal:fig17}
		\vspace{-0.4cm}
\end{figure}

In Fig.~\ref{SingleRISOAcal:fig17}, we investigate  classification accuracy  versus $K$ for different $L$ values. It is observed that the classification accuracy for  $L=1$ and $L=2$  initially increases monotonically with $K$ and then saturates. This behavior differs from the case where the over-the-air parameters are non-trainable, as shown in Fig.~\ref{SingleRISOAcal:fig15}. 
The is because, although the rank of the channel decreases with increasing $K$, the channel gain against noise improves, which benefits the training process. In contrast, in Fig.~\ref{SingleRISOAcal:fig15}, the reduction in channel rank directly leads to a larger emulation error, resulting in degraded classification performance.
Furthermore, it can be observed that the classification accuracy achieved with $L=5$ consistently outperforms that with $L=1$ and $L=2$. This improvement can be attributed to the fact that a larger 
$L$ yields a higher channel rank and increases the DoF available for optimizing the over-the-air parameters, thereby enhancing the classification performance.

\subsection{Fashion-MNIST dataset}
In this subsection, we present results with the Fashion-MNIST dataset.
 \begin{figure}[!t]
	\centerline{\includegraphics[width=3.2in]{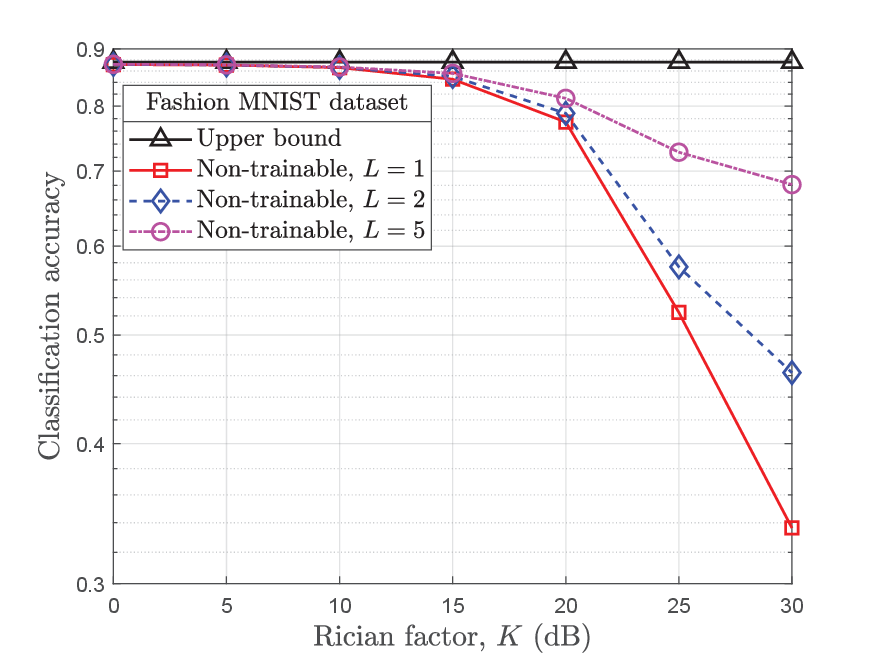}}
	\caption{Classification accuracy  versus $K$ for different  $L$ values under $M=100$ and $P_{\rm max} = 10~{\rm dB}$  base on Fashion-MNIST dataset. }  \label{Fashion:fig18}
\end{figure}
In Fig.~\ref{Fashion:fig18}, we consider the case where the over-the-air parameters are non-trainable and analyze the classification accuracy versus $K$ for different $L$ values. The ``Upper bound"  is obtained with the target NN  architecture.  It is observed that the curve trends in Fig.~\ref{Fashion:fig18} are similar to those in  Fig.~\ref{SingleRISOAcal:fig15}.
 \begin{figure}[!t]
	\centerline{\includegraphics[width=3.2in]{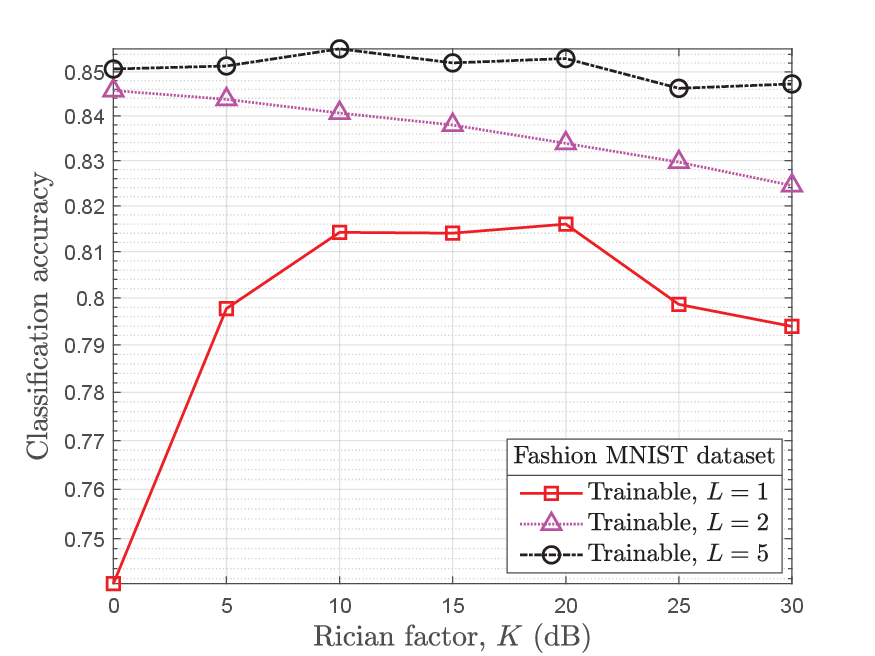}}
	\caption{Classification accuracy  versus $K$ for different $L$ values under $M=100$ and $P_{\rm max} = 10~{\rm dB}$ based on the Fashion-MNIST dataset.}  \label{Fashion:fig19}
\end{figure}

In Fig.~\ref{Fashion:fig19}, we consider the case with trainable parameters and study the classification accuracy versus $K$ for different $L$ values. It is observed that the curve trends in Fig.~\ref{Fashion:fig19} differ from those in  Fig.~\ref{SingleRISOAcal:fig17}. For $L=1$, the classification accuracy initially increases and then decreases with $K$ in Fig.~\ref{Fashion:fig19}, while it increases monotonically with $K$ in Fig.~\ref{SingleRISOAcal:fig17}. This difference arises because the Fashion-MNIST dataset is more complex, implying that a higher-rank FC layer is required to adequately represent the input features. In contrast, for the MNIST dataset in Fig.~\ref{SingleRISOAcal:fig17}, a simple low-rank FC layer is sufficient to capture the data features, and a higher channel gain against noise directly improves the performance. Moreover, it is observed that increasing $L$ leads to enhanced performance, as a higher channel rank can be achieved. This further demonstrates the advantage of adopting multi-RIS for AirFC computations.

\section{Conclusion}
In this paper, we studied RIS-aided MIMO  systems to engineer the ambient wireless propagation environment to emulate the  FC layer of a NN via analog OAC. To fully unveil the fundamental principle of RIS for analog OAC, we first investigated the case, in which the over-the-air system parameters are jointly optimized to minimize the mismatch between the OAC system and the target  FC layer.  
To address this non-convex optimization problem, a low-complexity alternating optimization algorithm was proposed, where semi-closed-form/closed-form solutions for all optimization variables are derived. Subsequently, we considered training of the system parameters using two distinct learning strategies, depending on the availability of CSI. While a centralized training approach is considered when the CSI is known, an alternative over-the-air training solution is also proposed that does not require CSI. We also extended our exploration to the multi-RIS scenario, leveraging the spatial diversity gain to enhance the classification accuracy for relatively simple classification problems. Simulation results demonstrated that AirFC via RIS can be realized in the analog domain, achieving satisfactory classification accuracy. Moreover, results indicated that utilizing additional multi-RISs leads to improved classification accuracy, particularly when the LoS component dominates the wireless channel. 

As future work, we will explore multi-hop RIS architectures and more complex network structures, e.g., convolutional layers, to further improve the  scalability of over-the-air analog neural networks.

\bibliographystyle{IEEEtran}
\bibliography{AirFCRIS}
\end{document}